\documentclass[preprint,prd,aps,floatfix,preprintnumbers,superscriptaddress,bibnotes,nofootinbib]{revtex4-1}

\usepackage{epsfig}
\usepackage{amsmath}
\usepackage{latexsym}
\usepackage[psamsfonts]{amssymb}
\usepackage{graphicx}
\usepackage{ulem}
\usepackage{longtable}
\usepackage{epstopdf}
\usepackage{bm}
\usepackage {tablefootnote} 
\usepackage{color}

\usepackage{comment}

\newcommand{\be}{\begin{equation}}
\newcommand{\ee}{\end{equation}}
\newcommand{\bea}{\begin{eqnarray}}
\newcommand{\eea}{\end{eqnarray}}
\newcommand{\bi}{\begin{itemize}}
\newcommand{\ei}{\end{itemize}}

\begin{document}
\preprint{KEK-TH-2837}

\title{
Spectral reconstruction from Euclidean lattice correlators through singular value decomposition
}

%
\author{Ryutaro Tsuji}
\email[E-mail: ]{rtsuji@post.kek.jp}
\affiliation{High Energy Accelerator Research Organization (KEK), Ibaraki 305-0801, Japan}
\author{Shoji Hashimoto}
\affiliation{High Energy Accelerator Research Organization (KEK), Ibaraki 305-0801, Japan}
\affiliation{School of High Energy Accelerator Science, The Graduate University for Advanced Studies, SOKENDAI, Ibaraki 305-0801, Japan}

\date{\today}
\begin{abstract}
Reconstructing spectral densities from Euclidean lattice correlators requires an inverse Laplace transform, which is inherently ill-conditioned when applied to numerical data with statistical uncertainties. The maximum amount of information that can be extracted from the imaginary-time dependence of correlators can be characterized by the singular value decomposition (SVD) of the kernel function $\exp(-\omega t)$ defined on discrete sets of imaginary times $t$ and energies $\omega$. The SVD provides orthogonal basis functions in both the $t$- and $\omega$-spaces, while the singular values determine the magnitude of their contributions to the correlators. By retaining only the components associated with the largest singular values, for which the correlator data remain statistically significant, one can reconstruct smeared spectral functions with controlled uncertainties. The systematic error arising from the truncation can also be bounded under reasonable assumptions. In the limit where the ranges of $t$ and $\omega$ become infinitely large and continuous, the SVD basis approaches the Mellin transform, allowing a representation of the smeared spectrum that is independent of the details of the lattice parameters.

\end{abstract}

\maketitle

 
\section{Introduction}
\label{sec:introduction}

In lattice Quantum Chromodynamics (QCD) calculations, the properties of hadrons are studied from correlation functions of local operators. 
By the K\"all\'en--Lehmann spectral representation, the Euclidean lattice correlator $C(t)$ as a function of Euclidean time $t$ is related to the spectral density $\rho(\omega)$ in the form of a Laplace transform
\begin{align}
    \label{eq:spec_rep}
    C(t) = \int_0^\infty d\omega\,\rho(\omega) e^{-\omega t}. 
\end{align}
To be concrete, we consider a zero-momentum projected two-point correlation function
\begin{align}
    \label{eq:twopt}
    C(t)
    & =
    \sum_{\bm{x}}
    \langle \Omega |
    O_{\chi}(t,\bm{x})
    \overline{O}_\chi(0,\bm{0})
    | \Omega \rangle,
\end{align}
where 
an interpolating operator $O_\chi(t,\bm{x})$ specifies the quantum number of states that contribute to the correlator.
The extraction of $\rho(\omega)$ from (\ref{eq:spec_rep}) requires an inverse Laplace transform, which poses a notoriously difficult problem \cite{MBertero_1985, MBertero_1988}, since the condition number of $\mathrm{e}^{-\omega t}$ when viewed as a matrix with discrete indices of $t$ and $\omega$ is huge.
In other words, the inverse problem admits infinitely many solutions for $\rho(\omega)$, that satisfy (\ref{eq:spec_rep}) within the statistical error. In order to circumvent this ill-posedness of the inverse problem, two approaches have been attempted.
Reconstruction using the maximum entropy method \cite{Nakahara:1999vy, Asakawa:2000tr}, the sparse modeling method \cite{Otsuki:2020spm, Itou:2020azb}, the Baysian reconstruction \cite{Burnier:2013nla,DelDebbio:2024lwm}, the Nevanlinna-Pick interpolation \cite{Bergamaschi:2023xzx,Fields:2025glg}, or the neural network and related \cite{Kades:2019wtd,Horak:2021syv,Buzzicotti:2023qdv, Aarts:2025gyp} is one class of such approaches, where additional assumptions are imposed when solving (\ref{eq:spec_rep}). They are subject to systematic bias depending on the assumptions.

Another approach is to redefine the problem so that one extracts some smeared spectral densities
\begin{align}
  \rho_{\sigma_\mathrm{smr}}(\omega)
  = \int^\infty_0 d\omega^\prime\,
  S_{\sigma_\mathrm{smr}}(\omega,\omega^\prime)\rho(\omega^\prime)
\end{align}
with a kernel $S_{\sigma_\mathrm{smr}}(\omega, \omega^\prime)$ that typically has a peak around $\omega$ with shoulders of some width $\sigma_\mathrm{smr}$. Sufficiently smooth kernels transform the problem to a better conditioned one, as we will discuss in this paper in detail.
To obtain the smeared spectrum, one tries to find an approximation of the form 
$S_{\sigma_\mathrm{smr}}(\omega,\omega^\prime)\simeq
c^{\sigma_\mathrm{smr}}_0(\omega)
+c^{\sigma_\mathrm{smr}}_1(\omega)e^{-\omega^\prime}
+\cdots
+c^{\sigma_\mathrm{smr}}_N(\omega) e^{-N\omega^\prime}$ 
truncated at $N$; the coefficients $c^{\sigma_\mathrm{smr}}_l(\omega)$ are given for each value of $\omega$ using methods such as the modified Backs-Gilbert (or HLT) approach \cite{Hansen:2019idp} or the Chebyshev approximation \cite{Bailas:2020qmv}.
The smeared spectrum $\rho_{\sigma_\mathrm{smr}}(\omega)$ can then be determined as
\begin{align}
    \rho_{\sigma_\mathrm{smr}}(\omega)
    \simeq
    \sum_{l=0}^{N}c^{\sigma_\mathrm{smr}}_l(\omega)\langle P_l(\hat{H})\rangle,
\end{align}
where $\langle P_l(\hat{H})\rangle$ is a matrix element of a polynomial of $e^{-\hat{H}}$ sandwiched by $\langle\Omega|O_\chi$ and $\bar{O}_\chi|\Omega\rangle$. Here, $\hat{H}$ represents a Hamiltonian whose explicit form is not necessary in the following discussion. The polynomial is given as $P_l(\hat{H})=e^{-l\hat{H}}$ for the HLT method and $P_l(\hat{H})=T^*_l(e^{-\hat{H}})$ for the shifted Chebyshev polynomials $T^*_l(x)$. The matrix element of each term $\langle e^{-t\hat{H}}\rangle$ is nothing but the correlator $C(t)$, and the smeared spectral function is obtained as long as the polynomial approximation is sufficiently precise. The idea has been applied to the $R$-ratio using the HLT method \cite{ExtendedTwistedMassCollaborationETMC:2022sta}.

The convergence of the approximation is the key issue; it is controlled by the coefficients $c^{\sigma_\mathrm{smr}}_l$ and the matrix elements $\langle P_l(\hat{H})\rangle$.
Typically, the coefficients  decrease rapidly for large $l$ when $\sigma_\mathrm{smr}$ is large, so the method works. But higher order terms are not necessarily suppressed in general, especially when the smearing width $\sigma_\mathrm{smr}$ is small. 
In addition, the matrix elements $\langle P_l(\hat{H})\rangle$ involve statistical errors, and the number of terms $N$ that can be included depends on the numerical precision of $C(t)$ at $t=N$.
Systematic errors due to ignored higher order terms need to be carefully examined. The relation between the Chebyshev approximation and the HLT approach is discussed in \cite{Barone:2023tbl}.

Either of these approaches, the inverse problem with some assumptions or the approximation of the smeared spectrum, has its own merits and problems. In order to clarify the limitation of these approaches, it is useful to step back and consider how much information is contained in the lattice correlator itself. The spectral representation (\ref{eq:spec_rep}) is based on the kernel $e^{-\omega t}$ integrated over $\omega$, with a weight $\rho(\omega)$, which means a sum over an indefinite number of states.
However, we do not know how many independent and numerically significant degrees of freedom are contained in the correlator itself. 

To this end, we consider a singular value decomposition (SVD) of the kernel $e^{-\omega t}$ taking a discrete set of values of $t$ and $\omega$ as indices:
\begin{equation}
\label{eq:SVD}
    e^{-\omega t} = \sum_l U_l(t)\sigma_l V_l(\omega),
\end{equation}
where $\sigma_l$ represents singular values labeled by an integer $l$. The  $U_l(t)$ and $V_l(\omega)$ are vectors associated with the singular value $l$, and they form orthonormal vectors, {\it i.e.} $\int_t U_l(t) U_{l'}(t)=\delta_{l,l'}$ and $\int_\omega V_l(\omega) V_{l'}(\omega)=\delta_{l,l'}$. (The integral over $t$ or $\omega$ is understood as a sum for the set of values of $t$ or $\omega$.) More precise definitions are presented in the next section.)
When listed in descending order, the singular values, which are defined to be real and positive, decrease exponentially. In realistic lattice calculations, the correlators have statistical error and the contributions of too small $\sigma_l$ terms can hardly be disentangled. Typically, only the terms of the largest $O(10)$ singular values are numerically significant, and that  represents the information that the lattice correlator data actually carry.

Using (\ref{eq:SVD}) and the orthogonality relations, the correlator $C(t)$ can be decomposed into the basis of $U_l(t)$ as 
\begin{equation}
    C(t)=\sum_l U_l(t)\sigma_l\rho_l, 
    \;\;\;
    \mathrm{where} 
    \;\;\;
    \rho_l=\int_\omega\rho(\omega)V_l(\omega).
\end{equation}
This construction makes it explicit that the $l$-th component contributes to the correlator with a weight of $\sigma_l$. 
Conversely, the spectral density is reconstructed from the correlator as 
\begin{equation}
    \rho(\omega)=\sum_l V_l(\omega) C_l/\sigma_l,
    \;\;\;
    \mathrm{where}
    \;\;\;
    C_l=\int_t U_l(t)C(t).
\end{equation}  
The contribution of the individual $l$ is enhanced by the factor of $1/\sigma_l$ when $\sigma_l$ is small, so that the statistical error in $C_l$ is amplified. This is precisely the source of the problem about the difficulty of the inverse Laplace transform. As we shall see in more detail in the following sections, the basis functions $V_l(\omega)$ oscillate more rapidly for larger $l$, so that fine details of the spectral function can only be reconstructed including large $l$ terms.
The smeared spectrum $\rho_{\sigma_\mathrm{smr}}(\omega)$ can be obtained more easily as
\begin{equation}
    \rho_{\sigma_\mathrm{smr}}(\omega)=\sum_l S_l(\omega) C_l/\sigma_l,
    \;\;\;
    \mathrm{where}
    \;\;\;
    S_l(\omega)=\int d\omega'\,S_{\sigma_\mathrm{smr}}(\omega,\omega')V_l(\omega').
\end{equation}  
When the smearing kernel $S_{\sigma_\mathrm{smr}}(\omega,\omega')$ does not contain a rapidly changing component probed by large $l$'s, $S_l(\omega)$ is suppressed and the enhanced error by the factor $1/\sigma_l$ is compensated.

This observation leads to a systematic strategy to represent the lattice correlator and the smearing kernel on the SVD basis. Reliable extraction of the smeared spectrum is limited to the cases where the statistical error of $C_l$ is sufficiently small when multiplied by $S_l(\omega)/\sigma_l$. This gives an explicit condition on what can be extracted from the lattice correlators.

The use of the SVD basis for the analysis of imaginary-time correlator has been introduced in condensed matter physics \cite{PhysRevLett.75.517,PhysRevB.96.035147,PhysRevB.98.035104} for the finite-temperature kernel $e^{-\omega t}/(1\pm e^{-\beta t})$. (See also an even earlier work \cite{TOMINAGA1986530}.) In this work, we focus on the zero-temperature kernel $e^{-\omega t}$ and enjoy a simplification as the range of $t$ and $\omega$ increases. That is, the SVD in our formulation approaches the Mellin transform recently introduced in \cite{Bruno:2024fqc}. Thus, we can define a set of basis functions independent of the lattice details, such as the Euclidean time extent and discretization.

This paper is organized as follows.
In Section~\ref{sec:singular_value_decomposition_of_laplace_kernel}, we formulate the SVD basis to approximate the spectral observables. 
Section~\ref{sec:mellin_tansform} discusses a relation between the Mellin transform and the SVD on a discrete lattice with finite time extent. It turns out that the SVD is identical to the Mellin transform in the limit of continuous and infinitely large time extent.
In Section~\ref{sec:modal_decomposition_of_spectral_function} we test our approach using a set of mock data, and Section~\ref{sec:numerical_application} discusses how the truncation error is suppressed for the smeared spectral density.
A summary and perspective are given in Section~\ref{sec:summary}.

\section{Singular value decomposition of Laplace kernel}
\label{sec:singular_value_decomposition_of_laplace_kernel}

Consider a lattice correlator $C(t)$ in $t\in[t_\mathrm{min}, t_\mathrm{max}]$, and assume that the spectrum $\rho(\omega)$ is limited in the range $\omega\in[\omega_\mathrm{min},\omega_\mathrm{max}]$.
For a discrete set of $t$ and $\omega$ defined as
$t_i=t_\mathrm{min}+i\Delta_t$ with $i=1,\dots,N_t$ and $\omega_j=\omega_\mathrm{min}+j\Delta_\omega$ with $j=1,\dots,N_\omega$, the spectral representation of correlator (\ref{eq:spec_rep}) can be viewed as a matrix equation: 
\begin{align}
    \label{eq:spec_dis}
    \tilde{C}(t_i) = \sum_{j=1}^{N_\omega} \tilde{L}_{ij}\tilde{\rho}(\omega_j),
\end{align}
where each quantity is normalized as
$\tilde{C}(t_i)\equiv\sqrt{\Delta_t}C(t_i)$,
$\tilde{\rho}(\omega_j)\equiv\sqrt{\Delta_\omega}\rho(\omega_j)$, 
and
$\tilde{L}_{ij} = \sqrt{\Delta_t} e^{-t_i\omega_j}\sqrt{\Delta_\omega}$.
Here, $\tilde{L}$ is a $N_t \times N_\omega$ rectangular matrix, which represents the Laplace kernel on a discrete grid.

The SVD of $\tilde{L}$ is written as
\begin{align}
    \tilde{L}_{ij}=\sum_{l=0}^{\mathrm{rank}(\tilde{L})-1} \tilde{U}_l(t_i)\sigma_l \tilde{V}_l(\omega_j)
\end{align}
with orthonormal vectors $\tilde{U}_l(t_i)$ and $\tilde{V}_l(\omega_j)$, each of which is associated with a singular value $\sigma_l$. To recover the Laplace kernel without the extra normalization one should use $U_l=\Delta_t^{-1/2}\tilde{U}_l$ and $V_l=\Delta_\omega^{-1/2}\tilde{V}_l$, thus $L_{ij}=\sum_{l=0}^{\mathrm{rank}(L)-1} U_l(t_i)\sigma_l V_l(\omega_j)$. The orthonormality of $\tilde{U}_l$ and $\tilde{V}_l$,
\begin{align}
    \sum_{i=1}^{N_t} \tilde{U}_l(t_i)\tilde{U}_{l^\prime}(t_i) = \delta_{l,l^\prime}
    \;\;\;
    \mathrm{and}
    \;\;\;
    \sum_{j=1}^{N_\omega} \tilde{V}_l(\omega_j)\tilde{V}_{l^\prime}(\omega_j) = \delta_{l,l^\prime},
\end{align}
holds with this normalization. 
We assume that the (real and positive) singular values are sorted in descending order ($\sigma_0\ge\sigma_1\ge\dots$). The vectors $\tilde{U}_l(t_i)$ and $\tilde{V}_l(\omega_j)$ represent oscillating functions of $t$ and $\omega$, respectively, and the number of nodes is given by $l$, as we shall see below. 

We can write (\ref{eq:spec_dis}) in the form
\begin{align}
    \tilde{C}(t_i) & = \sum_{l=0}^{\mathrm{rank}(\tilde{L})-1} \tilde{U}_l(t_i)\sigma_l \rho_l,
\end{align}
where $\rho_l$ represents the $l$-th mode of $\rho(\omega)$ in the SVD basis, that is
\begin{align}
    \label{eq:rhomode_omega}
    {\rho}_l & = \sum_{j=1}^{N_\omega} \tilde{V}_l(\omega_j) \tilde{\rho}(\omega_j).
\end{align}
When the correlator is given instead, $\rho_l$ is determined as
\begin{align}
    \label{eq:rhomode_t}
    {\rho}_l & =
    \frac{1}{\sigma_l} \sum_{i=1}^{N_t} \tilde{U}_l(t_i)\tilde{C}(t_i).
\end{align}
Since the singular value appears in the denominator, relatively small components of $\tilde{C}$ in the basis of $\tilde{U}_l$ are enhanced together with their noise.

%
%
\begin{figure*}[tb]
\centering
\includegraphics[width=0.6\textwidth,bb=0 0 425 305,clip]{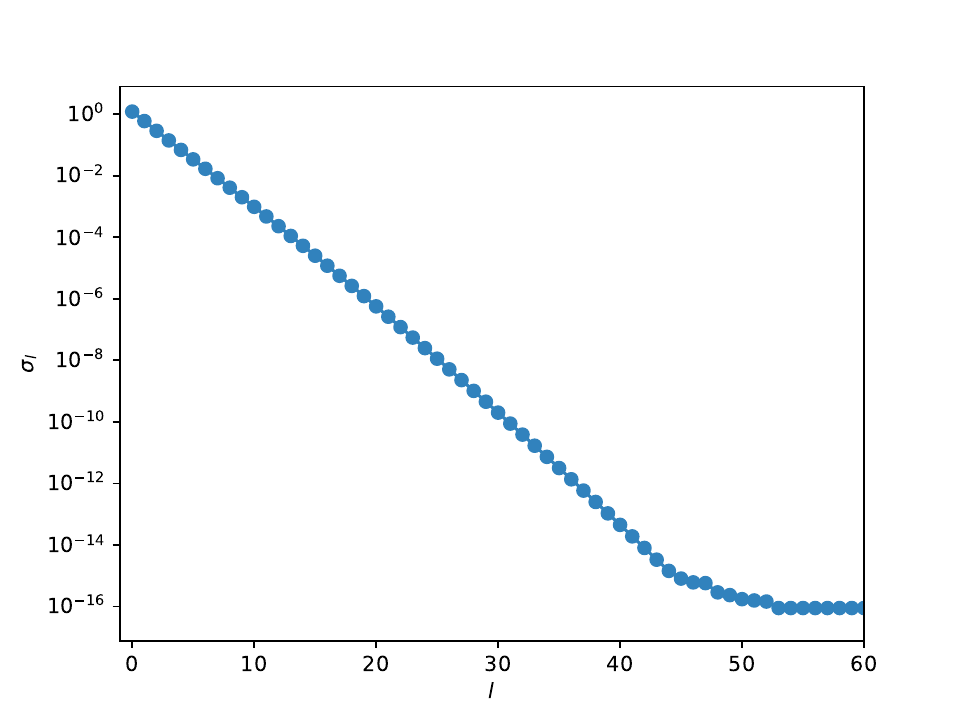}
\caption{
The singular values $\sigma_l$ of a square $\tilde{L}$ with $\Delta^t=\Delta^\omega=0.05$, $t_\mathrm{min}=\omega_\mathrm{min}=0.1$ and $t_\mathrm{max}=\omega_\mathrm{max}=50.1$. They are plotted in decreasing order of singular values.
}
\label{fig:symmetric_singularvalue}
\end{figure*}

In Fig.~\ref{fig:symmetric_singularvalue}, we plot the singular values $\sigma_l$ of $\tilde{L}$ with $\Delta_t=\Delta_\omega=0.05$, $t_\mathrm{min}=\omega_\mathrm{min}=0.1$ and $t_\mathrm{max}=\omega_\mathrm{max}=50.1$ We observe an exponential decrease of $\sigma_l$ with $l$. Around $l\sim$ 40--50, it flattens reaching the limit of double precision; even before, small $\sigma_l$'s are results of substantial cancellations.
Some of the corresponding vectors $\tilde{U}_l(t)$ and $\tilde{V}_l(\omega)$ are plotted in Figs.~\ref{fig:symmetric_rangedep_basisfunc_low}, \ref{fig:symmetric_rangedep_basisfunc_int}, \ref{fig:symmetric_rangedep_basisfunc_high} in Appendix~\ref{app:the_scaling_properties}. They show oscillations with frequency determined by $l$. 

We apply SVD to construct an approximation of $\rho(\omega)$ by truncating the series
\begin{align}
    \label{eq:rhorecon_omega}
    \tilde{\rho}^{(\mathrm{approx})}(\omega_j) & =
    \sum_{l=0}^{\mathrm{rank}(\tilde{L})-1} \tilde{V}_l(\omega_j){\rho}_l
\end{align}
at some order of $l$.
Even in the best case scenario, we must truncate at $l=N_\mathrm{tr}$, where $\sigma_l$ reaches the limit of double precision, because the value of $\rho_l$ beyond that is not reliable. In practice, we truncate where the statistical noise of $C(t)$ disrupts the signal of ${\rho}_l$. The truncation error can be bound as
\begin{align}
    \delta \tilde{\rho}^{(N_\mathrm{tr})}(\omega_k)
    & =
    |\tilde{\rho}(\omega_k) - \tilde{\rho}^{(N_\mathrm{tr})}(\omega_k) |
    =
    \left|
    \sum_{l=N_\mathrm{tr}}^{\mathrm{rank}(\tilde{L})-1}\tilde{V}_l(\omega_k){\rho}_l
    \right| 
    \nonumber\\
    & \le
    \sum_{l=N_\mathrm{tr}}^{\mathrm{rank}(\tilde{L})-1}
    |\tilde{V}_l(\omega_k)|\cdot|\rho_l|
    \label{eq:truncation_error}
\end{align}
using the triangle inequality. 
Since the basis $\tilde{V}_l(\omega)$ represents a function of $\omega$ that oscillates at a frequency of $O(l^{-1})$, the scaling of $\rho_l$ for large $l$ is given by the most rapid transition region of $\rho(\omega)$. For example, for a Breit-Wigner-type spectrum with a width $\Gamma$ the scaling would be like $\rho_l\sim e^{-\Gamma l}$. For a $\delta$-function spectrum, on the other hand, $\rho_l$ remains constant for large $l$. The estimate (\ref{eq:truncation_error}) gives the upper limit of the error according to the expected shape of the spectrum. Obviously, for the constant $\rho_l$ corresponding to the $\delta$-function-like spectrum, the error bound would be too loose to be useful.

There are physical observables that are written as an integral of $\rho(\omega)$ in some energy range. 
Consider a quantity $X$ in general,
\begin{align}
    \label{eq:spec_app}
    X & =
    \int^{\omega_\mathrm{max}}_{\omega_\mathrm{min}} d\omega\,
    K(\omega) \rho(\omega),
\end{align}
where $K(\omega)$ is a known kernel to define $X$. 
Using SVD, we can approximate $X$ as
\begin{align}
    \label{eq:spec_app_approx}
    X& \approx 
    \sum_{l=0}^{N_\mathrm{tr}-1} \rho_l
    \left[
    \sum_{j=1}^{N_\omega} \tilde{K}(\omega_j) \tilde{V}_l(\omega_j)
    \right] 
    = 
    \sum_{l=0}^{N_\mathrm{tr}-1} \rho_l \tilde{K}_l,
\end{align}
where the series is truncated at $l=N_\mathrm{tr}-1$, and the coefficients are constructed as
$\tilde{K}_l = \sum_{j=1}^{N_\omega} \tilde{K}(\omega_j) \tilde{V}_l(\omega_j)$ and $\tilde{K}(\omega_j)=\sqrt{\Delta_\omega}K(\omega_j)$.
The component of the spectrum $\rho_l$ is obtained from the correlator, while $\tilde{K}_l$ is calculated simply once the SVD of $e^{-\omega t}$ is given. It corresponds to an approximation of the kernel function $K(\omega)$ in terms of the basis functions $\tilde{V}_l(\omega)$, {\it i.e.} 
$K(\omega)\simeq\sum_{l=0}^{N} \tilde{K}_l \tilde{V}_l(\omega)$.
The truncation induces systematic uncertainty, which is estimated by
\begin{align}
    \label{eq:truncation_error_app}
    \delta X^{(N_\mathrm{tr})}
    \le
    \sum_{l=N_\mathrm{tr}}^{\mathrm{rank}(\tilde{L})-1}
    |\tilde{K}_l| |\rho_l|.
\end{align}
This is suppressed more when $K(\omega)$ is a slowly varying function, because then $|\tilde{K}_l|$ decreases exponentially.

This construction is common for any choice of kernel function $K(\omega)$. The lattice correlator provides $\rho_l$, while the kernel part $\tilde{K}_l$ is built independently of the lattice data. Therefore, for example, the smeared spectral function around any value of $\omega$ with any choice of smearing is constructed from the same set of $\rho_l$'s. The precision that can be obtained is determined by the rate of exponential decrease of $|K_l|$ for that particular choice, as well as that of $|\rho_l|$.

This observation implies an inherent limitation in the reconstruction of $\rho(\omega)$ (or its smeared version) from $C(t)$. The SVD of $e^{-\omega t}$ extracts the relevant modes of $C(t)$ on a certain orthogonal basis. Since its singular value decreases exponentially for large $l$, typically only 10-20 modes are numerically significant due to the noise in the data. The reconstruction of $X$ discussed above is possible only when the kernel $K(\omega)$ is well approximated within this limited space. Extrapolation to vanishing smearing width, {\it e.g.} to obtain the unsmeared spectrum, necessarily involves the uncertainty from the smaller singular-value modes, since the number of available modes is unchanged. This limitation must be reflected in the systematic error for the smeared spectrum or other quantities obtained by the energy integral.

\section{Connection between SVD and the Mellin transform}
\label{sec:mellin_tansform}

For continuous and infinitely large time duration, {\it i.e.} $t\in[0,\infty)$ and $\omega\in[0,\infty)$, the decomposition of the Laplace kernel $e^{-\omega t}$ is given using the Mellin transform \cite{Bruno:2024fqc}, which is based on mathematical literature \cite{bellman1984laplace, JGMcWhirter_1978, doi:10.1137/060657273}. 
Here, we briefly review this formalism and then discuss the relation to the SVD introduced in the previous section.

Introducing a new variable $s\in\mathbb{R}$, the Laplace kernel is decomposed as
\begin{align}
    \label{eq:kernel_expansion_cont}
    e^{-\omega t}
    & = \int^\infty_{-\infty} ds\, u_s(\omega) \lambda_s^*u_s(t),
\end{align}
where the eigenvalues are given as $\lambda_s=\Gamma(\frac{1}{2}+is)$ using the Euler Gamma function $\Gamma(z)$.
The Mellin basis function is $u_s(t)=e^{is\ln(t)}/\sqrt{2\pi t}$. Although $\lambda_s$, $u_s(\omega)$ and $u_s(t)$ are complex-valued functions, the integral (\ref{eq:kernel_expansion_cont}) is real; we can write an equivariant formula with real eigenvalues and basis functions as
\begin{align}
    \label{eq:kernel_expansion_cont_real}
    e^{-\omega t}
    & =
    \int^\infty_0 ds\, |\lambda_s|
    \left(
    u^{+(\theta)}_s(\omega) u^{+(\theta)}_s(t)
    -
    u^{-(\theta)}_s(\omega) u^{-(\theta)}_s(t)
    \right)
\end{align}
with 
\begin{align}
    \label{eq:real_eigenvalue}
    |\lambda_s| & = 
    \sqrt{\frac{\pi}{\cosh(\pi s)}}.
\end{align}
Absorbing the complex phase of $\lambda_s$, the even $(+)$ and odd $(-)$ basis functions are defined as
\begin{align}
    u^{+(\theta)}_s(x) =
    \frac{\cos\left(s\ln(x)-\frac{\theta(s)}{2}\right)}{\sqrt{\pi x}},
    \;\;
    u^{-(\theta)}_s(x) =
    \frac{\sin\left(s\ln(x)-\frac{\theta(s)}{2}\right)}{\sqrt{\pi x}},
\end{align}
where $x=\omega$ or $t$, and $\theta(s) = \mathrm{arg}(\lambda_s)$. 
In terms of a logarithmic variable $X=\ln x$, the Mellin basis essentially corresponds to $\{\cos(sX), \sin(sX)\}$, which is analogous to the Fourier-transform.

In order to establish the connection with the SVD discussed in the previous section, we consider the decomposition (\ref{eq:kernel_expansion_cont_real}) in a finite interval defined by $t_\mathrm{min}=\omega_\mathrm{min}\equiv x_\mathrm{min}$ and $t_\mathrm{max}=\omega_\mathrm{max}\equiv x_\mathrm{max}$.
First, the finite interval would quantize $s$, which should correspond to the index $l$ in the SVD of $e^{-\omega t}$. In analogy to the Fourier transform, one expects a sequence of $s$ to be $s_l\sim l\Delta s$ with a spacing $\Delta s\sim \pi/\ln(x_\mathrm{max}/x_\mathrm{min})$. Then, it leads to an ansatz $|\lambda_{s_l}|\approx \sigma_l$, where $\sigma_l$ is the sigular value obtained by the SVD of $e^{-\omega t}$. For large $x_\mathrm{max}/x_\mathrm{min}$, $s_l$ would fill their gap continuously as $\Delta s\to 0$.
We approximate $u_s^{\pm(\theta)}(x)$ in the interval as
\begin{align}
    \label{eq:realeven_basis_functions_disc}
    u^{+(\theta)}_{s_l}(x)
    &\approx
    \sqrt{\frac{2}{x\ln(x_\mathrm{max}/x_\mathrm{min})}} 
    \cos\left(
    s\ln \frac{x}{\sqrt{x_\mathrm{max}x_\mathrm{min}}}
    -\frac{\theta(s_l)}{2}+l\pi
    \right),\\
    \label{eq:realodd_basis_functions_disc}
    u^{-(\theta)}_{s_l}(x)
    &\approx
    -\sqrt{\frac{2}{x\ln(x_\mathrm{max}/x_\mathrm{min})}} 
    \cos\left(
    s\ln \frac{x}{\sqrt{x_\mathrm{max}x_\mathrm{min}}}
    -\frac{\theta(s_l)}{2}+l\pi + \frac{\pi}{2}
    \right)
\end{align}
in analogy to the Fourier-transform. It has the form 
$(A_l/\sqrt{x})\cos\left(B_l \ln(x)+C_l\right)$.
With the numerical data of $U_l\propto \sqrt{1/x}\cos(B_l\ln(x)+C_l)$, we evaluate $B_l$ and $C_l$, which correspond to $|\lambda_s|$ and $\theta(s)$ as
\begin{align}
    \label{eq:fit_correspondence}
    B_l = s_l,\;\; 
    C_l =
    \left\{
    \begin{matrix}
    -\frac{\theta(s_l)}{2}, & \mathrm{even}\ l \\
    -\frac{\theta(s_l)}{2} + \frac{\pi}{2}, & \mathrm{odd}\ l
    \end{matrix}
    \right.
\end{align}
modulo $2\pi$ and a constant shift of the phase.

Assuming that $U_l$ is approximated by (\ref{eq:realeven_basis_functions_disc}) and (\ref{eq:realodd_basis_functions_disc}), we evaluate $s_l$ and $\theta(s_l)$ by numerically fitting $U_l$. (Here, we are referring the unrenormalized $U_l$, not $\tilde{U}_l$.) Then, $|\lambda_{s_l}|$ is evaluated using (\ref{eq:real_eigenvalue}), so that we obtain $\lambda_{s_l}=|\lambda_{s_l}|e^{i\theta(s_l)}$ from $U_l$ for a given $l$. Thus, we can compare $|\lambda_s|$ with the corresponding singular value $\sigma_l$ to confirm the correspondence.

%
%
\begin{figure*}[tbp]
\centering
\includegraphics[width=0.49\textwidth,bb=0 0 425 325,clip]{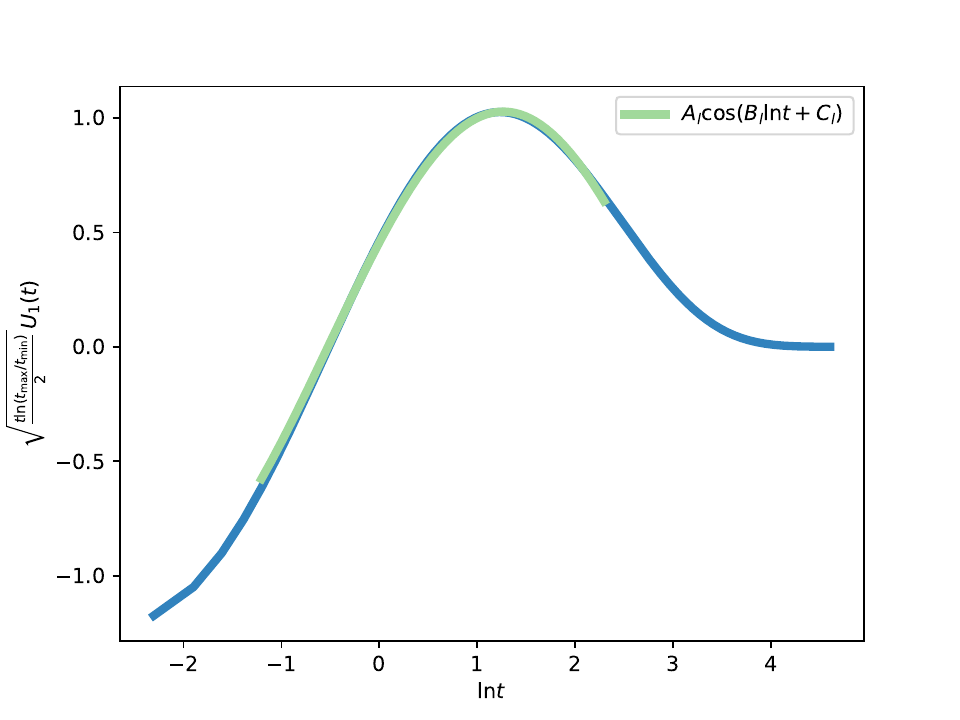}
\includegraphics[width=0.49\textwidth,bb=0 0 425 325,clip]{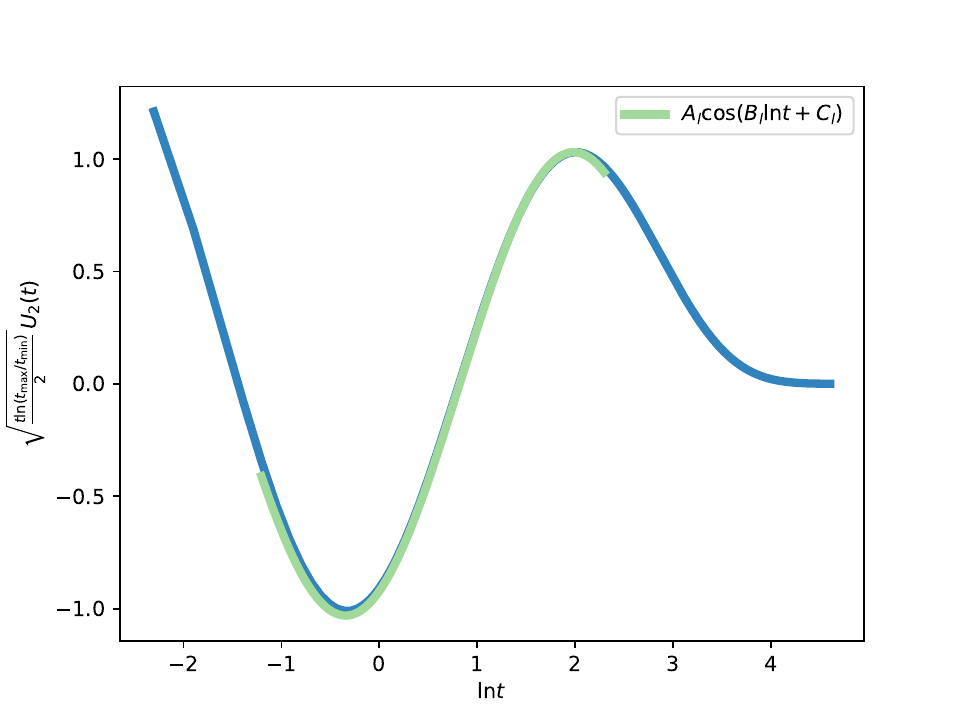}
\includegraphics[width=0.49\textwidth,bb=0 0 425 305,clip]{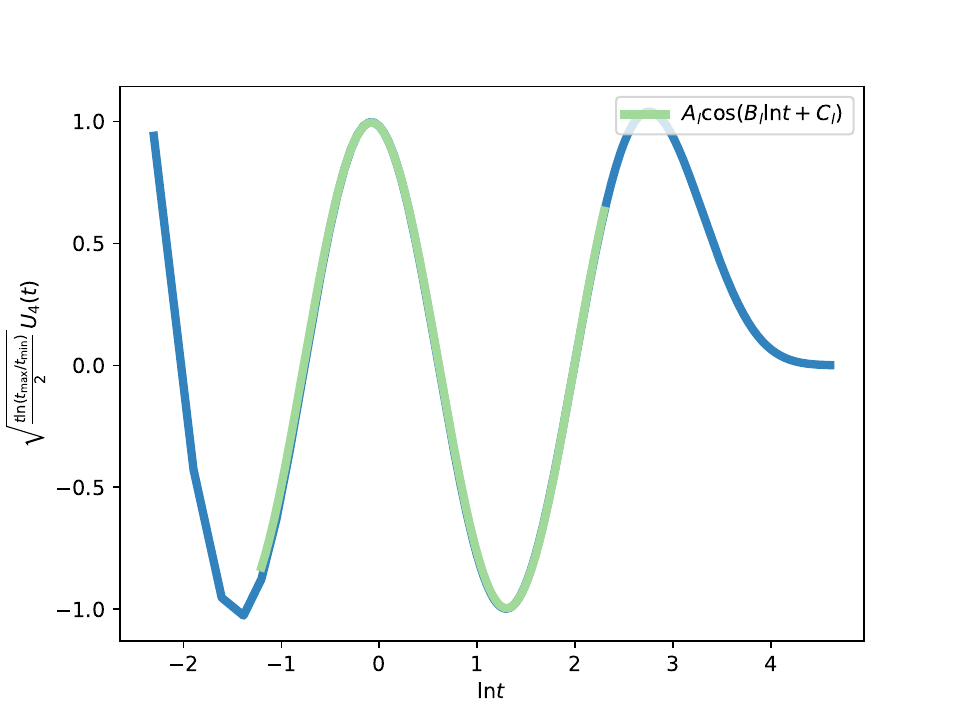}
\includegraphics[width=0.49\textwidth,bb=0 0 425 305,clip]{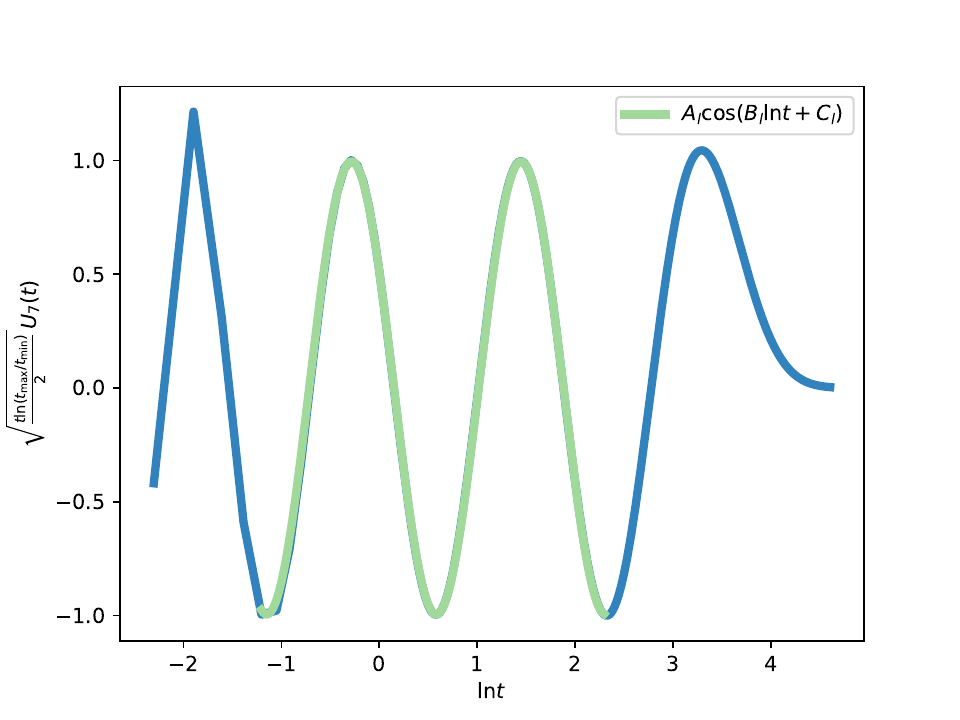}
\includegraphics[width=0.49\textwidth,bb=0 0 425 305,clip]{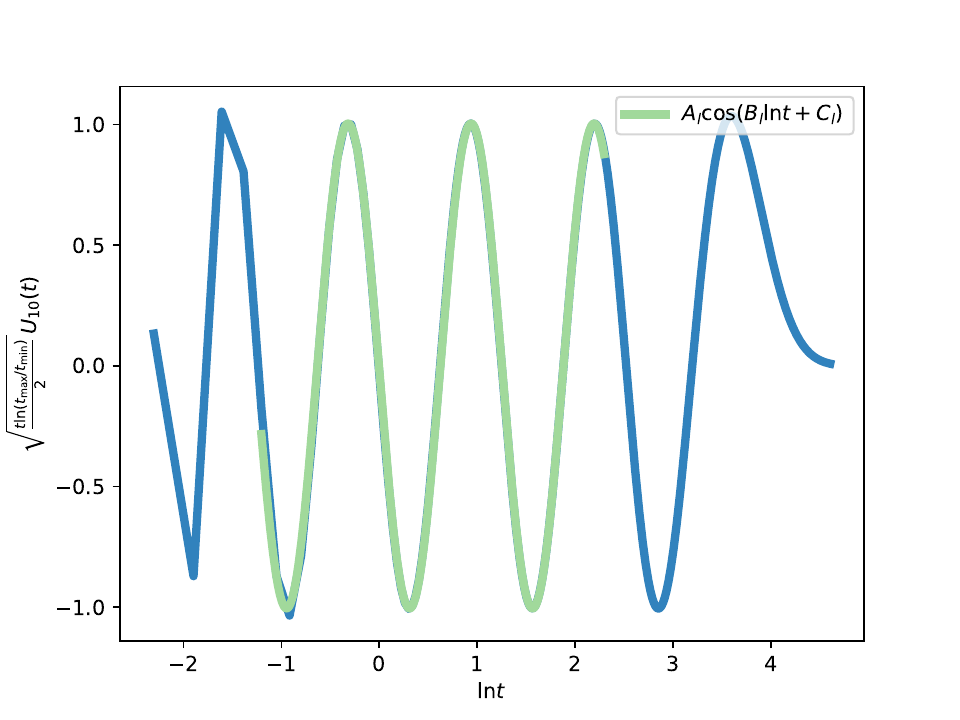}
\includegraphics[width=0.49\textwidth,bb=0 0 425 305,clip]{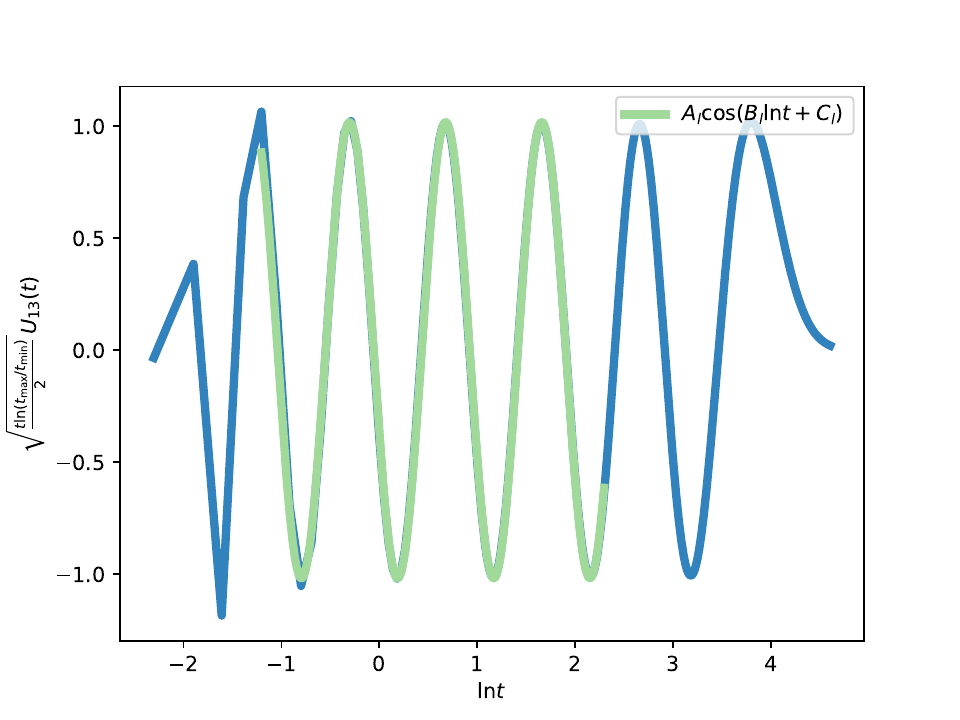}
\caption{
Rescaled basis functions of $U_l(t)$ from the SVD (blue) and the fit results (green). Each plot is obtained from $l=1$ (upper left), $2$ (upper right), $4$ (middle left), $7$ (middle right), $10$ (lower left), and $13$ (lower right).
}
\label{fig:mellinlimit_ufunc_ldep}
\end{figure*}

Let us first investigate how well $U_l$ from the SVD is approximated by (\ref{eq:realeven_basis_functions_disc}) and (\ref{eq:realodd_basis_functions_disc}).
We take $t_\mathrm{min}=\omega_\mathrm{min}=0.1$, $t_\mathrm{max}=\omega_\mathrm{max}=100.1$, $\Delta_t=\Delta_\omega\equiv\Delta=0.05$, and $N_t=N_\omega$. Fig.~\ref{fig:mellinlimit_ufunc_ldep} shows the rescaled basis functions $\sqrt{t\ln(t_\mathrm{max}/t_\mathrm{min})/2}\times U_l(t)$ (blue) and the fit results with $(A_l/\sqrt{t})\cos\left(B_l \ln(t)+C_l\right)$ (green) for several $l$'s. We take $-1.2\le \ln(t)\le+2.3$ to fit the data. 
Similarly to the Fourier transform, the $l$-th basis function has $l$ nodes. We observe that the rescaled basis function in the bulk can be well described by the cosine functions (\ref{eq:realeven_basis_functions_disc}) or (\ref{eq:realodd_basis_functions_disc}), while finding slight deviation for the edges due to boundary effects. 

%
%
\begin{figure*}[tb]
\centering
\includegraphics[width=0.49\textwidth,bb=0 0 425 325,clip]{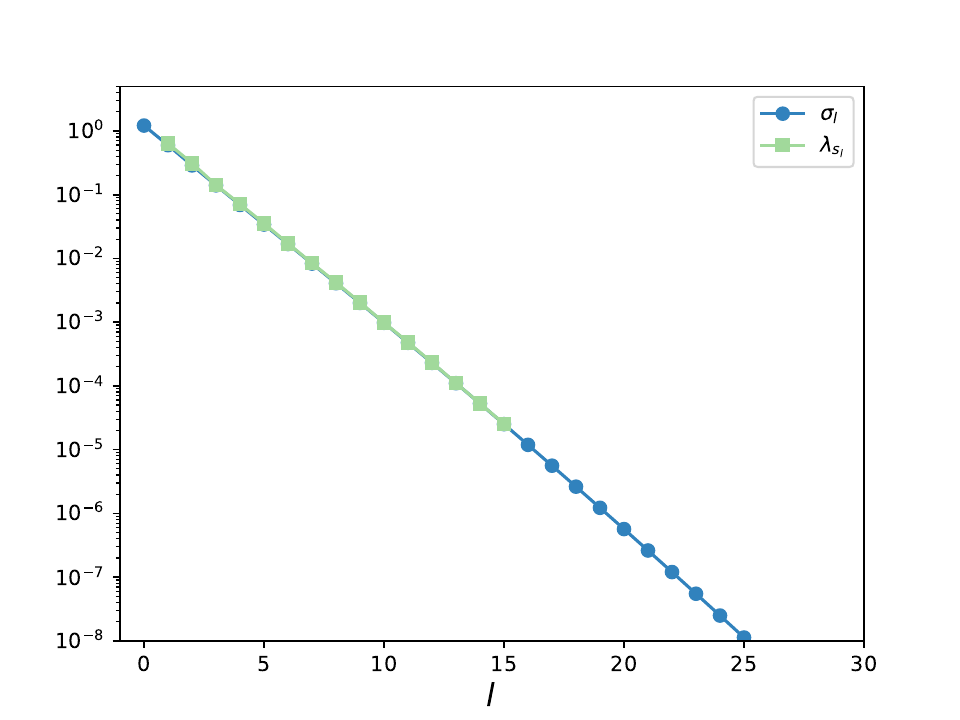}
\includegraphics[width=0.49\textwidth,bb=0 0 425 325,clip]{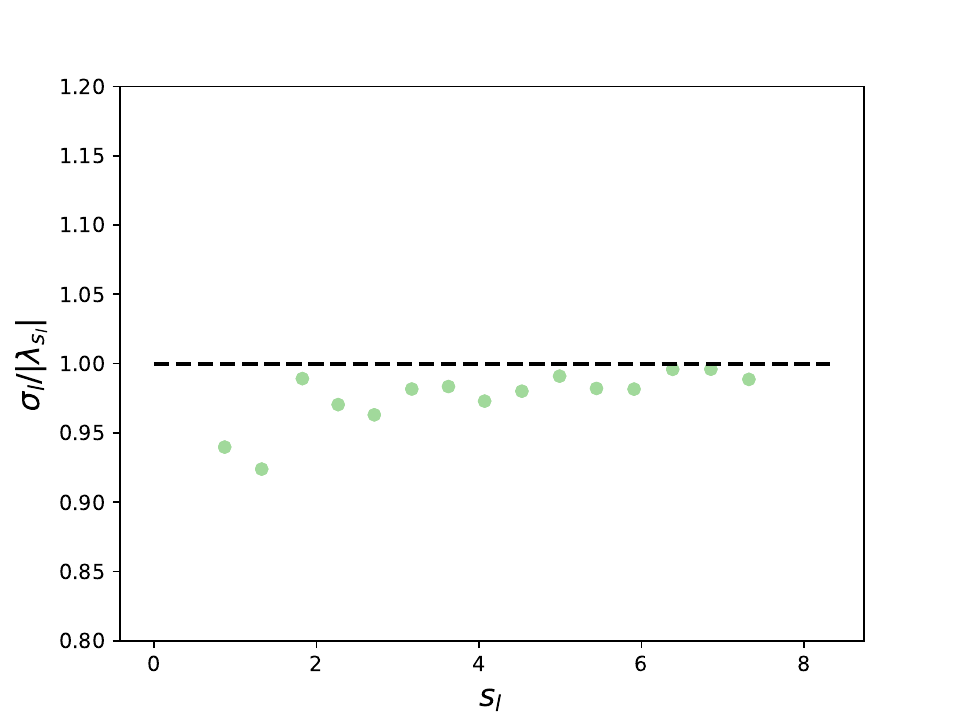}
\caption{
Left: $\sigma_l$ obtained from SVD of $e^{-\omega t}$ (blue circle) and $|\lambda_{s_l}|$ (green square) evaluated by (\ref{eq:real_eigenvalue}) with $B_l$ for $l\in[1,15]$. Right: their ratio $\sigma_l/|\lambda_{s_l}|$.
}
\label{fig:mellinlimit_singularvalue}
\end{figure*}

We reconstruct $|\lambda_{s_l}|$ from $B_l=s_l$ as previously described and compare it with $\sigma_l$ in  Fig.~\ref{fig:mellinlimit_singularvalue}. 
We observe that $\sigma_l$ given by the SVD numerically reproduces $|\lambda_{s_l}|$ (see Fig.~\ref{fig:mellinlimit_singularvalue} (left)). 
In the right panel, we plot the ratio $\sigma_l/|\lambda_{s_l}|$, with which we confirm that the agreement is within 10\% for $l\in[1,15]$ for the choice of parameters given above.

%
%
\begin{figure*}[tb]
\centering
\includegraphics[width=0.49\textwidth,bb=0 0 425 305,clip]{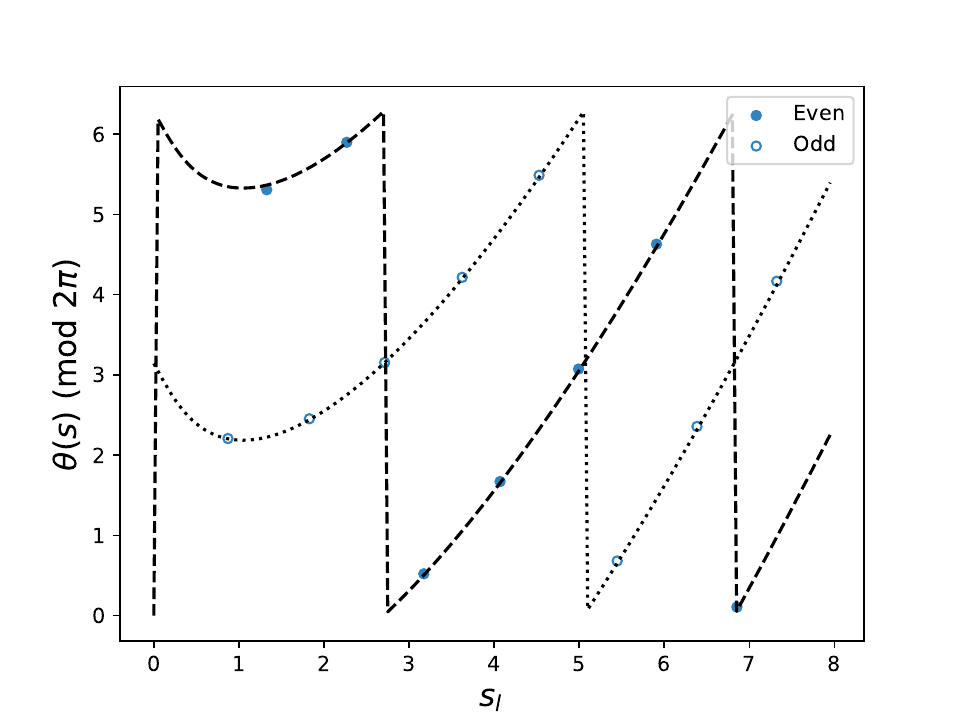}
\caption{
$\theta(s_l)\ (\mathrm{mod}\ 2\pi)$ from $C_l$. The $\theta(s_l)$ with even (odd) $l$ is denoted by filled (open) symbols. The dashed line presents $\mathrm{arg}[\Gamma(\frac{1}{2}+is)] \ (\mathrm{mod}\ 2\pi)$, while the dotted shows $\mathrm{arg}[\Gamma(\frac{1}{2}+is)]+\pi \ (\mathrm{mod}\ 2\pi)$.
}
\label{fig:mellinlimit_singularvalue_angle}
\end{figure*}

Fig.~\ref{fig:mellinlimit_singularvalue_angle} presents the phase $\theta(s_l)\ (\mathrm{mod}\ 2\pi)$ evaluated by (\ref{eq:fit_correspondence}) for $l\in[1,15]$. Each data point numerically reproduces the relation $\mathrm{arg}(\lambda_s)=\mathrm{arg}[\Gamma(\frac{1}{2}+is)]$ given by the Mellin transform. Even and odd $l$'s behave differently as indicated by (\ref{eq:realeven_basis_functions_disc}) and (\ref{eq:realodd_basis_functions_disc}). Recall that $\lambda_s$ in (\ref{eq:kernel_expansion_cont}) has a complex phase, which is absorbed into the phase shift of the basis functions in (\ref{eq:kernel_expansion_cont_real}).

%
%
\begin{figure*}[tb]
\centering
\includegraphics[width=0.49\textwidth,bb=0 0 425 325,clip]{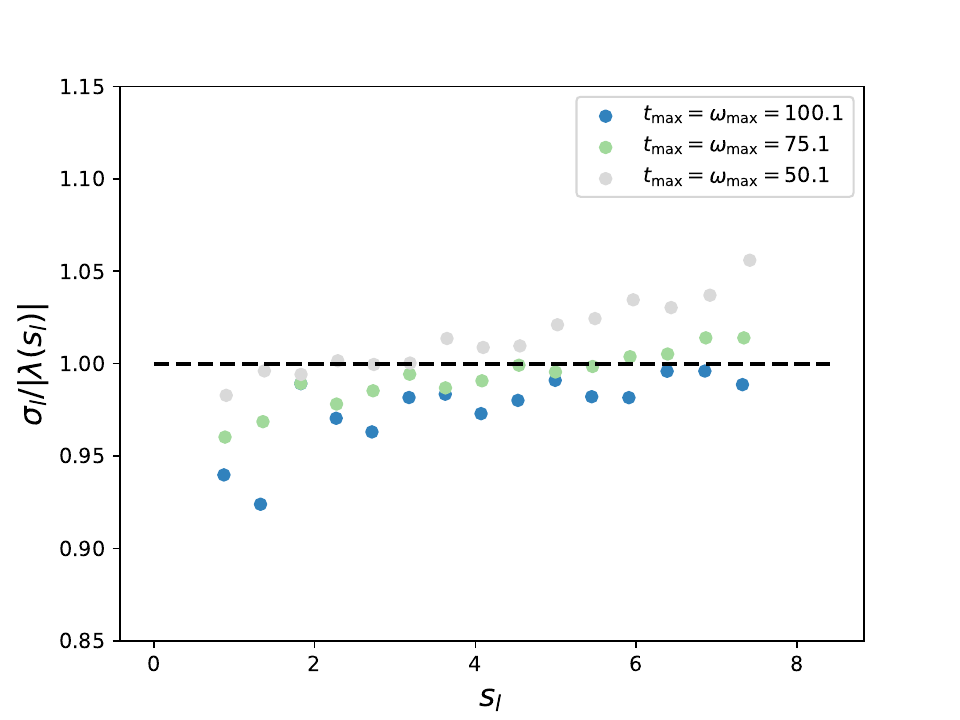}
\includegraphics[width=0.49\textwidth,bb=0 0 425 325,clip]{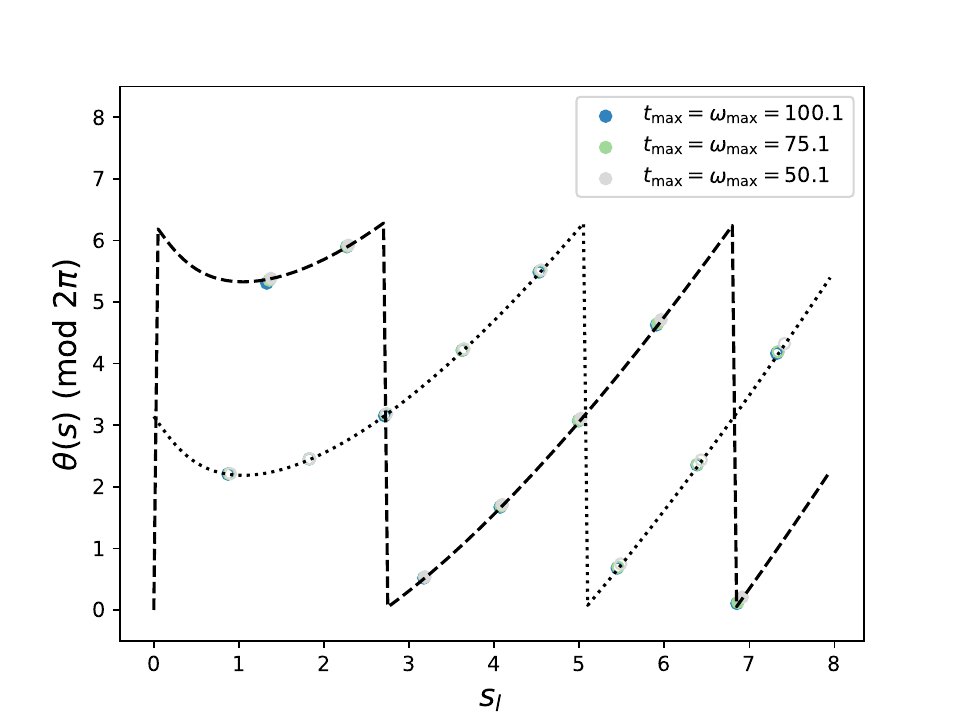}
\caption{
Left: $\sigma_l/|\lambda_{s_l}|$ with $t_\mathrm{max}=\omega_\mathrm{max}=100.1$ (blue), $75.1$ (green), and $50.1$ (gray) for $l\in[1,15]$. Right: $\theta(s_l)\ (\mathrm{mod}\ 2\pi)$. Even (odd) $l$'s are denoted by filled (open) symbols. We take $t_\mathrm{min}=\omega_\mathrm{min}=0.1$, and $\Delta_t=\Delta_\omega\equiv\Delta=0.05$.
}
\label{fig:mellin_rangedep_singularvalue}
\end{figure*}

Overall, we confirm that the singular values and the orthogonal basis generated by the SVD reproduce the corresponding objects in the Mellin transform. Now, we investigate how the limit to the continuous and infinite interval is approached.

We first consider the scaling of $\lambda_{s_l}$ with respect to various  $t_\mathrm{max}=\omega_\mathrm{max}$. We chose $t$ and $\omega$ symmetrically; different choices are discussed later. The singular values (Fig.~\ref{fig:symmetric_rangedep_singularvalue}) and the basis functions (Figs.~\ref{fig:symmetric_rangedep_basisfunc_low}, \ref{fig:symmetric_rangedep_basisfunc_int}, \ref{fig:symmetric_rangedep_basisfunc_high}) are plotted for $t_\mathrm{max}=\omega_\mathrm{max}$ = 50.1, 75.1, 100.1 in Appendix~\ref{app:the_scaling_properties}.
Here, Fig.~\ref{fig:mellin_rangedep_singularvalue} shows the ratio $\sigma_l/|\lambda_{s_l}|$ in the left panel, while $\theta(s_l)$ is shown in the right panel. The value of $s_l$ for each $l$, which is the horizontal axis, is nearly unchanged for different $t_\mathrm{max}=\omega_\mathrm{max}$ for the low-lying modes plotted here. This is consistent with the observation that the low-lying eigenmodes plotted in Fig.~\ref{fig:symmetric_rangedep_basisfunc_low} coincide for different $t_\mathrm{max}=\omega_\mathrm{max}$ and nearly vanish for $t\sim$ 50. More detailed comparison of the eigenmodes with different $t_\mathrm{max}=\omega_\mathrm{max}$ is found in Appendix~\ref{app:the_scaling_properties} (Fig.~\ref{fig:symmetric_rangedep_basisfunc_low}), where one can see that the low-modes are rarely affected by the large $t$ region probably because the kernel $e^{-\omega t}$ is highly suppressed. 
In Fig.~\ref{fig:mellin_rangedep_singularvalue} we observe a good agreement with the Mellin transform for both $\sigma_l/|\lambda_{s_l}|$ and $\theta(s_l)$. Some deviation of a few \% is found for $\sigma_l/|\lambda_{s_l}|$, but this is attributed to the range of $t_\mathrm{min}=\omega_\mathrm{min}$ as discussed below.

%
%
\begin{figure*}[tb]
\centering
\includegraphics[width=0.49\textwidth,bb=0 0 425 325,clip]{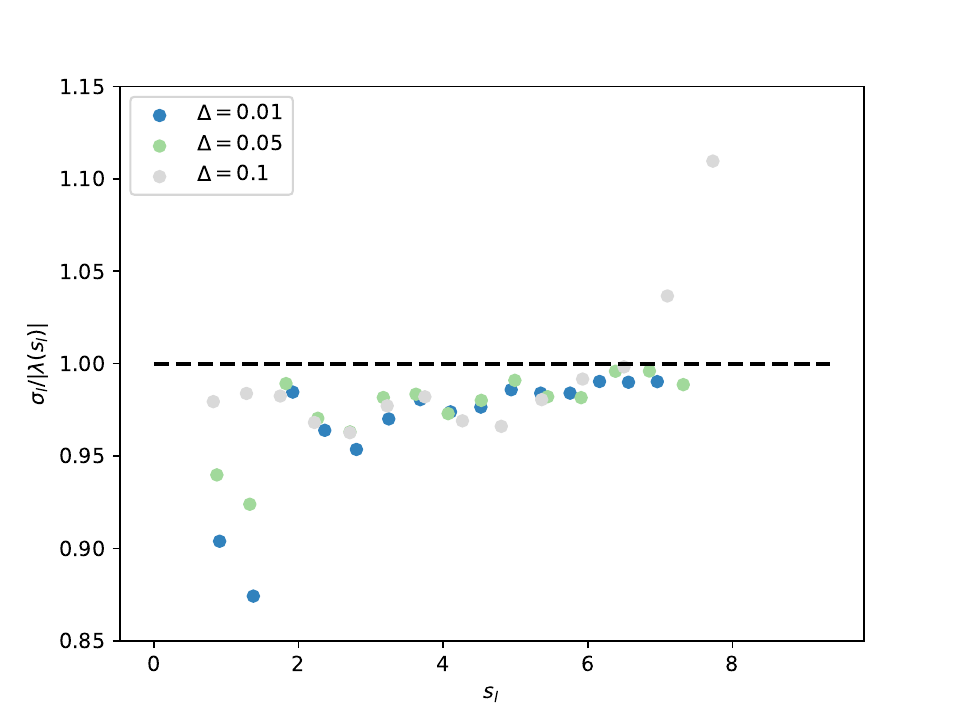}
\includegraphics[width=0.49\textwidth,bb=0 0 425 325,clip]{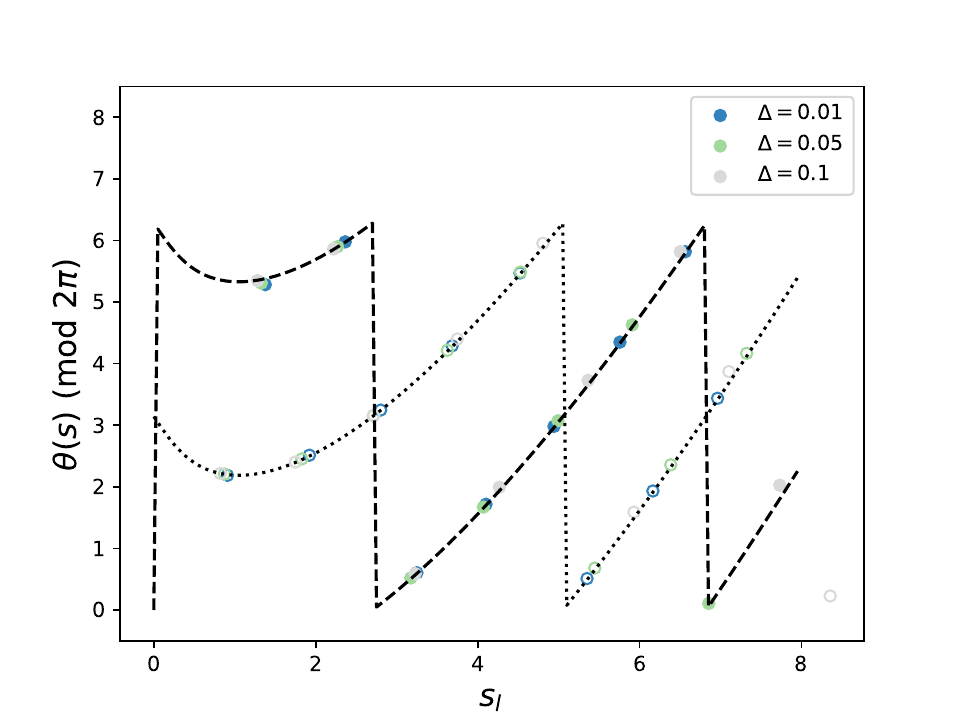}
\caption{
Left: $\sigma_l/|\lambda_{s_l}|$ obtained from $\Delta=0.01$ (blue), $0.05$ (green), and $0.1$ (gray) for $l\in[1,15]$. Right: $\theta(s_l)\ (\mathrm{mod}\ 2\pi)$. Even (odd) $l$'s are plotted by filled (open) symbols. We take $t_\mathrm{min}=\omega_\mathrm{min}=0.1$, and $t_\mathrm{max}=\omega_\mathrm{max}=100.1$.
}
\label{fig:mellin_griddep_singularvalue}
\end{figure*}

Next, we examine the dependence on $\Delta\equiv\Delta_t=\Delta_\omega$. We fix $t_\mathrm{max}=\omega_\mathrm{max}$ = 100.1. 
In Fig.~\ref{fig:mellin_griddep_singularvalue}, we plot $\sigma_l/|\lambda_{s_l}|$ ($l\in[1,15]$) for $\Delta=0.01$ (blue), $0.05$ (green) and $0.1$ (gray) on the left panel, while $\theta(s_l)$ is shown on the right panel. 
The agreement with the expectation from the Mellin transform is confirmed.

\begin{figure*}[tb]
\centering
\includegraphics[width=0.49\textwidth,bb=0 0 425 325,clip]{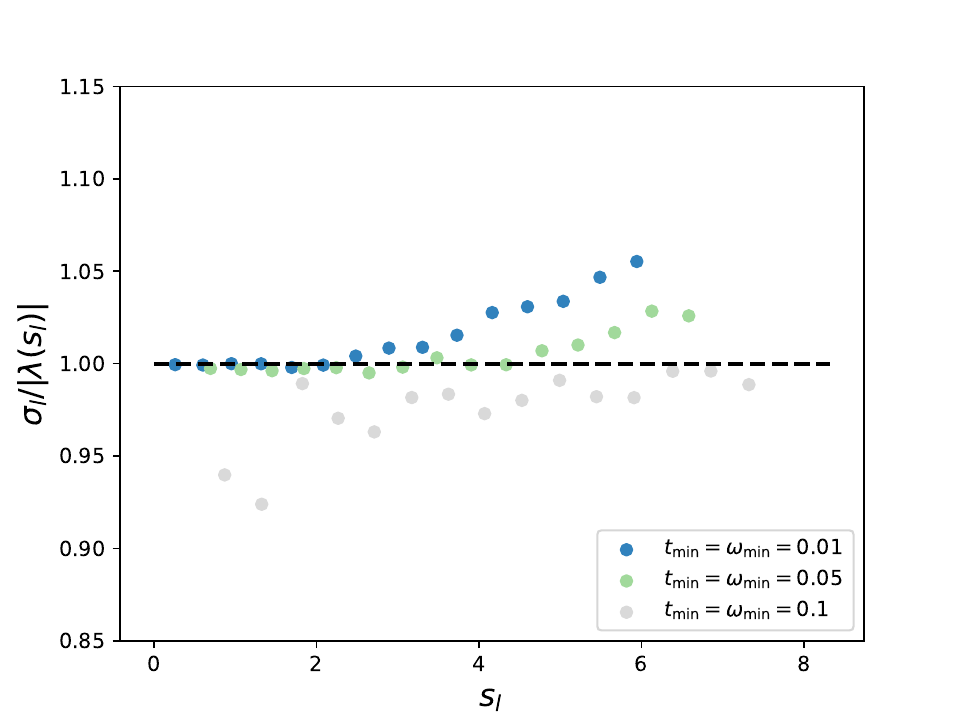}
\includegraphics[width=0.49\textwidth,bb=0 0 425 325,clip]{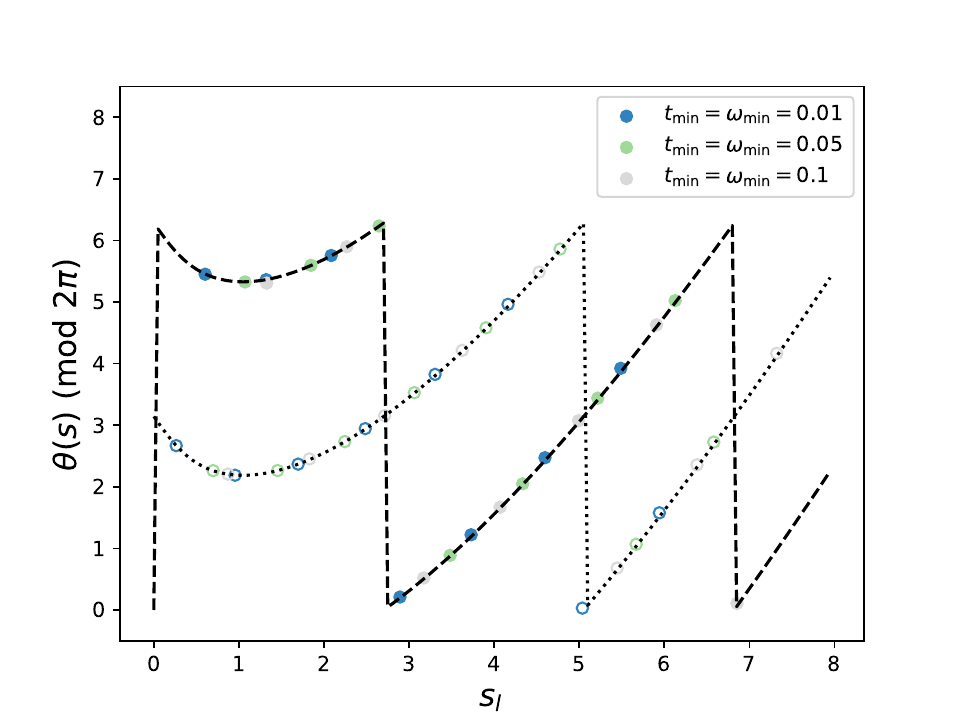}
\caption{
Left: $\sigma_l/|\lambda_{s_l}|$ obtained from $t_\mathrm{min}=0.01$ (blue), $0.05$ (green), and $0.1$ (gray) for $l\in[1,15]$. Right:
$\theta(s_l)\ (\mathrm{mod}\ 2\pi)$. Even (odd) $l$'s are denoted by filled (open) symbols. We take  $t_\mathrm{max}=\omega_\mathrm{max}=100.1$, and $\Delta_t=\Delta_\omega\equiv\Delta=0.05$.
}
\label{fig:mellin_Mindep_singularvalue}
\end{figure*}

Finally, Fig.~\ref{fig:mellin_Mindep_singularvalue} shows a similar comparison with $t_\mathrm{min}=\omega_\mathrm{min}=0.01$, $0.05$ and $0.1$. The $t_\mathrm{max}=\omega_\mathrm{max}$ is set to 100.1. Again, the expectation from the Mellin transform is well reproduced. 
With smaller $t_\mathrm{min}=\omega_\mathrm{min}$ one can access the smaller values of $s$ and the separation between neighboring $l$'s becomes narrower. It would eventually reproduce the continuous $s$ as $t_\mathrm{min}=\omega_\mathrm{min}\to 0$. Remarkably, we observe that the small $t_\mathrm{min}=\omega_\mathrm{min}$ improves the agreement of $\sigma_l/|\lambda(s_l)|$ with the Mellin transform. With $t_\mathrm{min}=\omega_\mathrm{min}=0.01$, the deviation is much smaller than 1\% for low-lying modes $s_l\lesssim 2$.

Through these observations, we have confirmed that $\sigma_l$, $U_l(t_i)$ and $V_l(\omega_j)$ approach the formula (\ref{eq:kernel_expansion_cont_real}) in the limits of $\Delta\to0$ and $t_\mathrm{min}=\omega_\mathrm{min}\to 0$, $t_\mathrm{max}=\omega_\mathrm{max}\to\infty$.
With finite but small $t_\mathrm{min}=\omega_\mathrm{min} \lesssim 0.01$, low modes below $s_l\lesssim 2$ reproduce the relation $\sigma=\lambda_s=\sqrt{\pi/\cosh(\pi s)}$ rather precisely. This suffices in practical analysis because $\rho_l$ can be determined well only for low modes due to the statistical noise of $C(t)$.

When applying for the lattice data of $C(t)$, we would set $\Delta_t=1$ in the unit of lattice spacing, while $\Delta_\omega$ should be kept small. Such asymmetric cases can be related to the symmetric one by a rescaling. Starting from the symmetric choice, $\bar{t}_\mathrm{min}=\bar{\omega}_\mathrm{min}=\bar\Delta\ll 1$, we apply a scale transformation $t=\lambda\bar{t}$, $\omega=\lambda^{-1}\bar{\omega}$ with $\lambda=1/\bar{\Delta}$, so that $\Delta_t=1$ and $\Delta_\omega=\Delta^2$.
With this transformation, $\tilde{U}_l(t)=\frac{1}{\sqrt{\lambda}}\tilde{U}_l(\lambda\bar{t})$ and $\tilde{V}_l(\omega)=\sqrt{\lambda}\tilde{V}_l(\lambda^{-1}\bar{\omega})$, 
where the prefactor $\sqrt\lambda^{(-1)}$ is assigned to adjust the normalization of the integral over $t$ or $\omega$.
By construction, to reach small energy scales $\omega\ll 1$, one needs a correlator at long distances $t\gg 1$.
One can even apply SVD for a rectangular matrix $N_t\neq N_\omega$. In either case, $\sigma_l$, $U_l(t_i)$ and $V_l(\omega_j)$ should numerically reproduce the Mellin transform as long as $\Delta_t\Delta_\omega$ is sufficiently small and $(t_\mathrm{max}/t_\mathrm{min})(\omega_\mathrm{max}/\omega_\mathrm{min})$ is large enough.

\section{SVD mode decomposition of spectral density}
\label{sec:modal_decomposition_of_spectral_function}

The properties of the singular values and associated basis vectors described in the previous section suggest a new strategy to approximate smeared spectral functions as outlined in the Introduction. Since it has a definite limit for the continuous and infinite range of $t$ and $\omega$, it can be defined independently of the lattice spacing and the temporal extent of the lattice. In the limit of infinite range, it goes back to the Mellin transform proposed in \cite{Bruno:2024fqc}.

We test the method using mock data derived from a given $\rho(\omega)$ in $[0.1, 5.1]$ with $\Delta_\omega=0.0005$. We define a Breit-Wigner-type true function 
\begin{align}
    \label{eq:breit_wigner}
    \rho^{(\mathrm{true})}(\omega;\omega_0)
    =
    \frac{1}{\pi}
    \frac{2\Gamma}{(\omega-\omega_0)^2+\Gamma^2},
\end{align}
A mock Euclidean correlator $C(t)$ is obtained by (\ref{eq:spec_rep}).

\begin{figure*}[tb]
\centering
\includegraphics[width=0.49\textwidth,bb=0 0 425 325,clip]{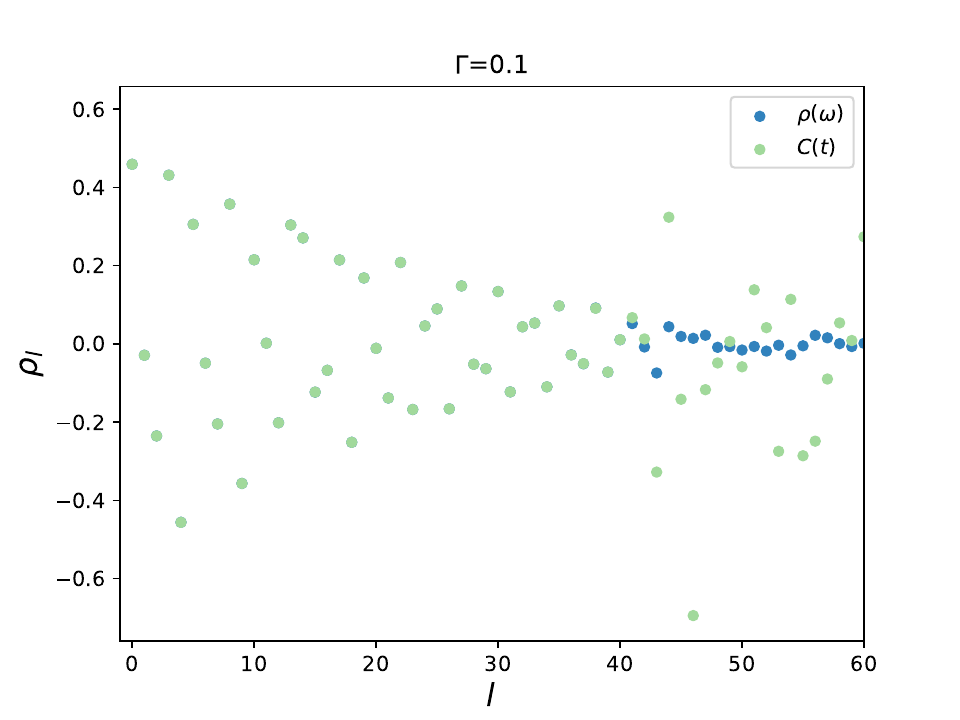}
\includegraphics[width=0.49\textwidth,bb=0 0 425 325,clip]{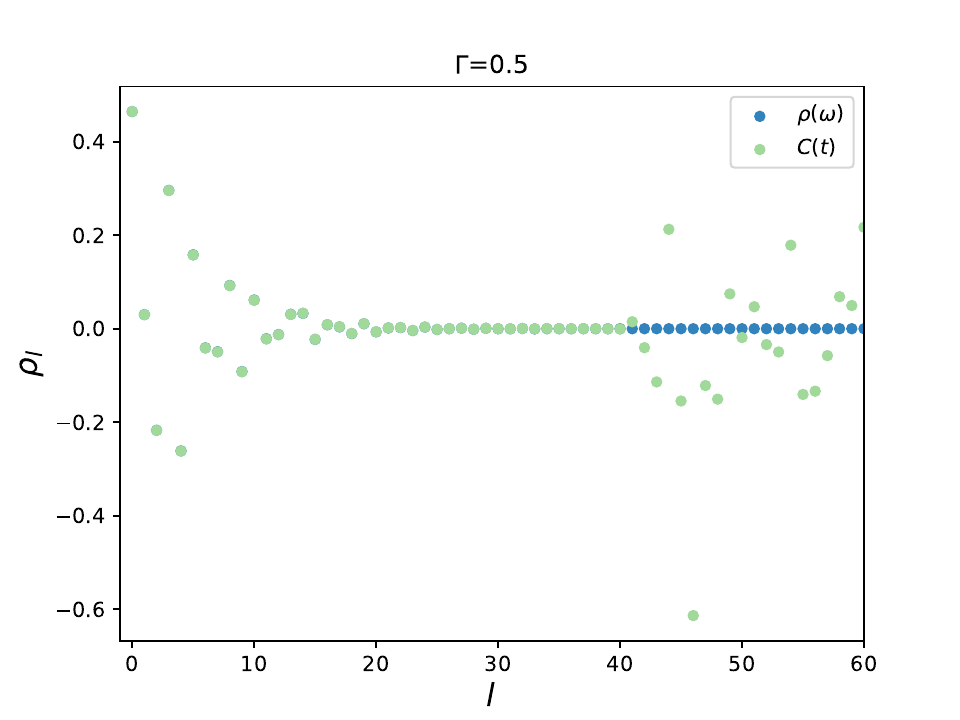}
\caption{
SVD components $\rho_l$ obtained from $\rho(\omega)$ (blue circle) and the $C(t)$ (green square) for $\Gamma=0.1$ (left) and $0.5$ (right).
}
\label{fig:mockbreitwigner_modal_decomposition_gammadep}
\end{figure*}

Fig.~\ref{fig:mockbreitwigner_modal_decomposition_gammadep} presents $\rho_l$ constructed directly from $\rho(\omega)$ using the basis vectors of SVD as (\ref{eq:rhomode_omega}) or through $C(t)$ as (\ref{eq:rhomode_t}). We take $\omega_0=1.0$, and the plots are for $\Gamma=0.1$ (left) and 0.5 (right).
We observe that the SVD mode components $\rho_l$ from two approaches agree precisely to $l\sim 40$. Recall that $\sigma_l$ of $\tilde{L}$ reaches the limit of double precision for $l\ge40$, and the significant deviation beyond $l\sim 40$ is attributed to the limitation of numerical precision. When the width $\Gamma$ is large, $\Gamma=0.5$, $\rho_l$ rapidly decreases and nearly vanishes around $l\sim 20$, while the corresponding plot for $\Gamma=0.1$ does not show the saturation at $l\sim 40$.

%
%

\begin{figure*}[tb]
\centering
\includegraphics[width=0.49\textwidth,bb=0 0 425 325,clip]{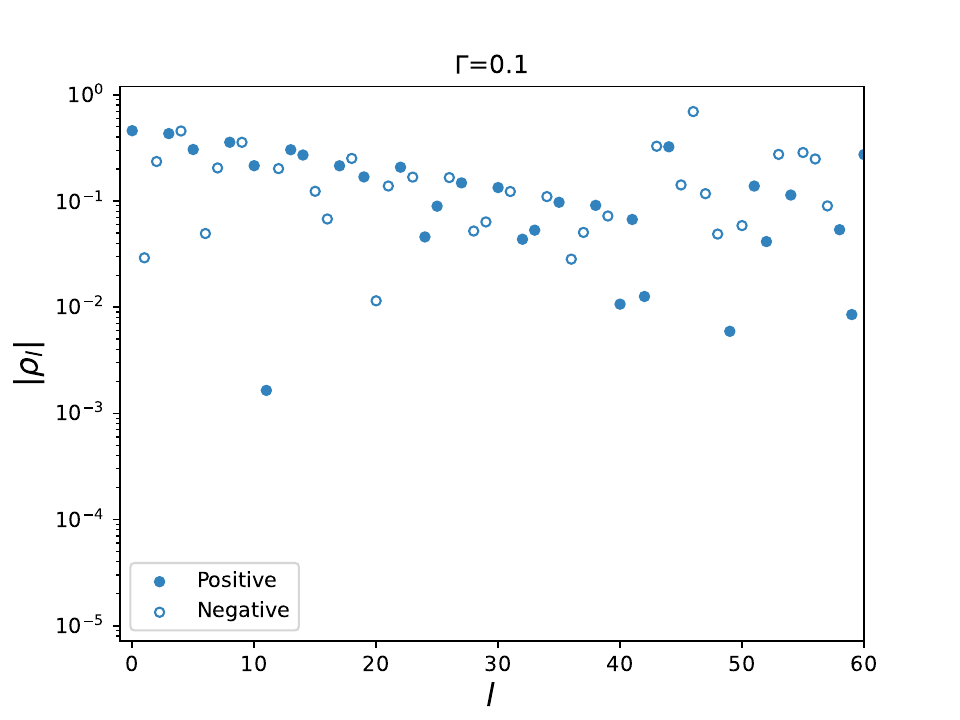}
\includegraphics[width=0.49\textwidth,bb=0 0 425 325,clip]{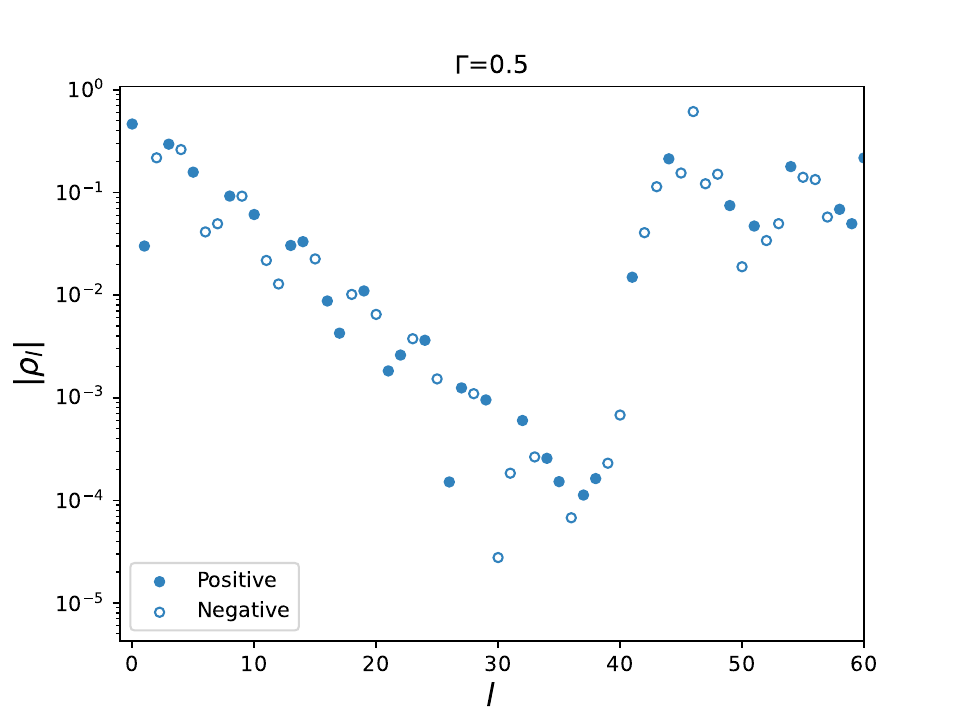}
\caption{
$|\rho_l|$ reconstructed from $C(t)$ for $\Gamma=0.1$ (left) and $0.5$ (right). The sign of $\rho_l>0$ is indicated by the symbols, {\it i.e.} positive (negetive) by filled (open) symbols.
}
\label{fig:mockbreitwigner_rho_size_gammadep}
\end{figure*}

The decrease in $|\rho_l|$ is typically exponential, as shown in Fig.~\ref{fig:mockbreitwigner_rho_size_gammadep}, which plots a scaling of $|\rho_l|$ for $\Gamma=0.5$ (left) and $0.1$ (right). Due to the limitation of double precision, the numbers beyond $l\gtrsim 39$ are unreliable. However, with the results $l\lesssim 38$ one can draw an envelope such that $|\rho_l|\leq A_\rho e^{-B_\rho l}$. The exponential slope $B_\rho$ is steeper for larger $\Gamma$. Assuming that the exponential decrease observed in these plots continues beyond $l=39$, we can estimate the truncation error by (\ref{eq:truncation_error}). 

%
%
%

%
%
\begin{figure*}[tb]
\centering
\includegraphics[width=0.49\textwidth,bb=0 0 425 325,clip]{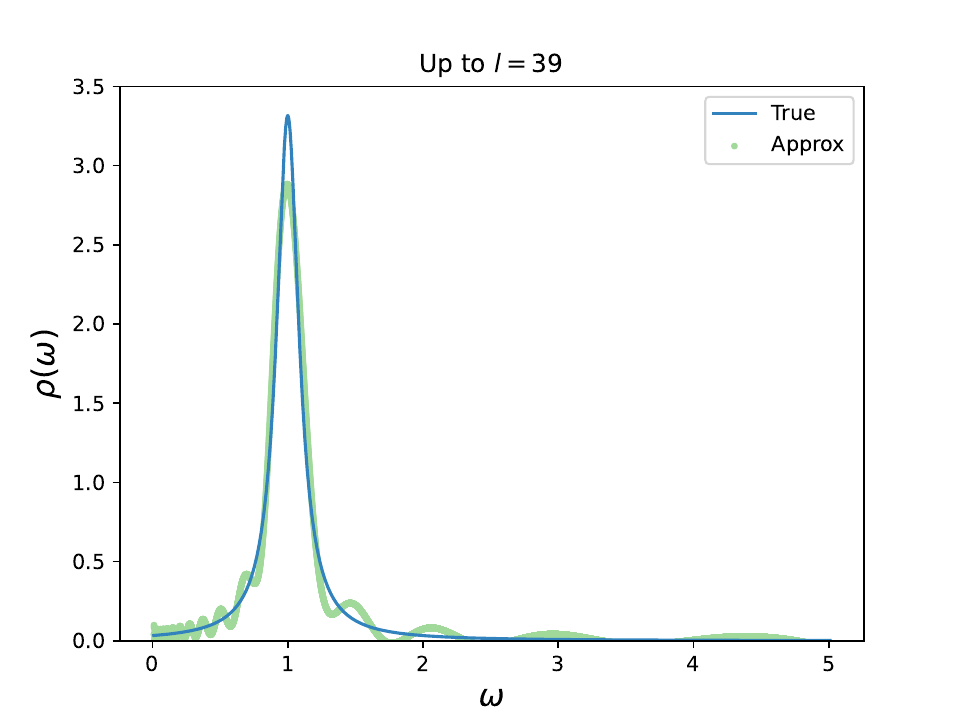}
\includegraphics[width=0.49\textwidth,bb=0 0 425 325,clip]{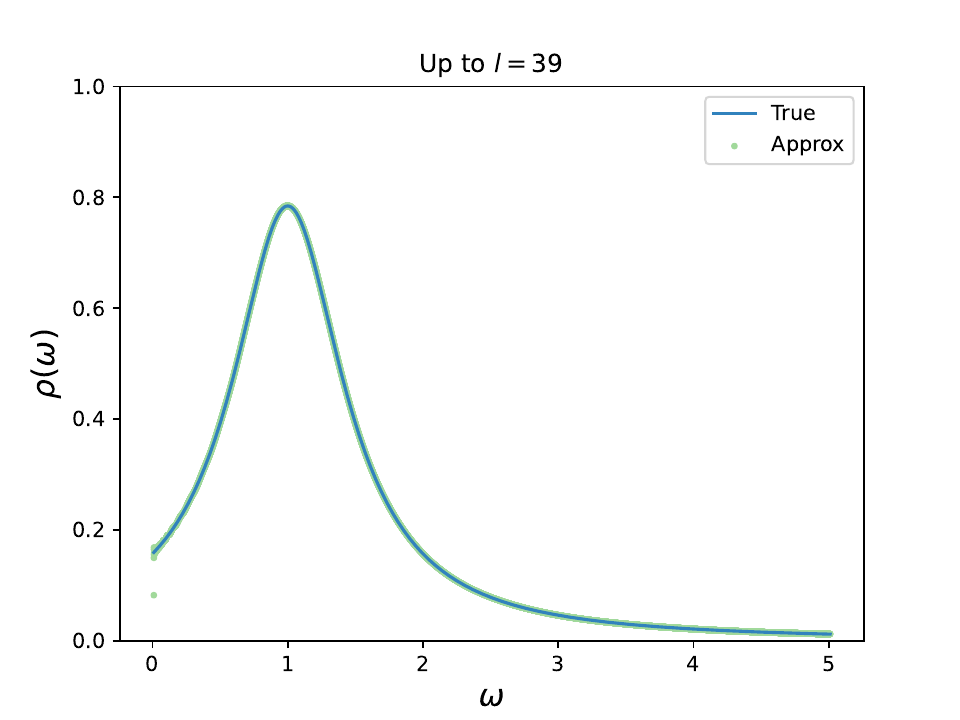}
\caption{
A comparison between the true $\rho(\omega)$ (blue line) and the approximated one with (\ref{eq:rhorecon_omega}) (green dot) using up to $l=39$.
Each plots is obtained with $\Gamma=0.1$ (left) and $0.5$ (right).
}
\label{fig:mockbreitwigner_reconstruction_gammadep}
\end{figure*}

Fig.~\ref{fig:mockbreitwigner_reconstruction_gammadep} shows a comparison between $\rho^{(\mathrm{true})}(\omega)$ (blue line) and $\rho^{(\mathrm{approx})}(\omega)$ with (\ref{eq:rhorecon_omega}) (green dot) using up to $l=39$. We confirm that $\rho^{(\mathrm{true})}(\omega)$ for $\Gamma=0.5$ is precisely approximated, while an oscillation is observed for $\Gamma=0.1$. This indicates that $\tilde{V}_l(\omega_j)$ in (\ref{eq:rhorecon_omega}) up to $l=39$ does not have enough resolution to precisely trace the peak structure.

%
%
\begin{figure*}[tb]
\centering
\includegraphics[width=0.49\textwidth,bb=0 0 425 325,clip]{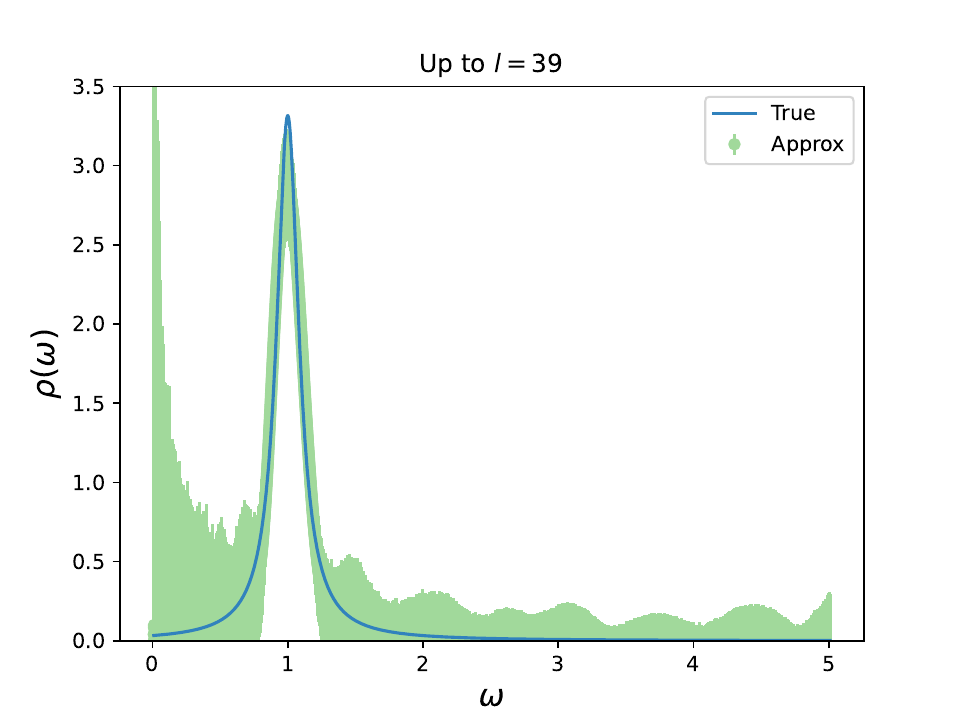}
\includegraphics[width=0.49\textwidth,bb=0 0 425 325,clip]{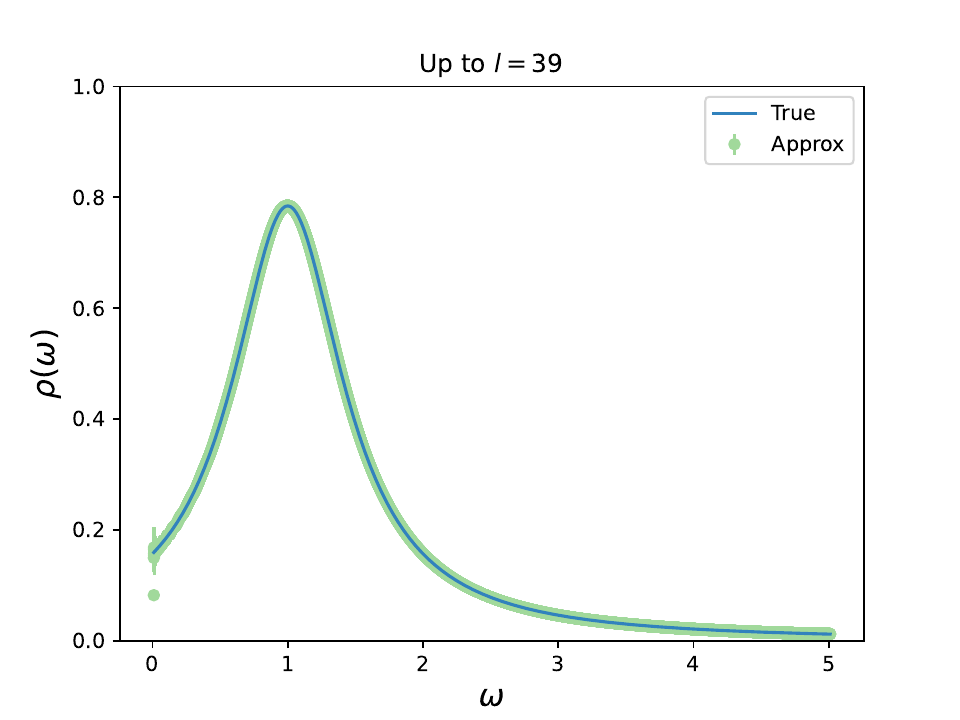}
\caption{
$\rho(\omega)$ (green dot) with an estimate of truncation error using up to $l=39$, compared with the true function (blue line). 
The data with $\Gamma=0.1$ (left) and $0.5$ (right).
}
\label{fig:mockbreitwigner_reconstructionerr_gammadep}
\end{figure*}

With the assumed scaling of $|\rho_l|$, the truncation error in $\rho^{(\mathrm{approx})}(\omega)$ can be estimated by (\ref{eq:truncation_error}). We evaluate the constants $A_\rho$ and $B_\rho$ that satisfy $|\rho_l|\lesssim A_\rho e^{-B_\rho l}$ using the data for $l\in[20,40]$.  Fig.~\ref{fig:mockbreitwigner_reconstructionerr_gammadep} shows plots of $\rho^{(\mathrm{approx})}(\omega)$ together with an estimate of the truncation error.
The truncation error for $\Gamma=0.5$ is not significant; $\rho^{(\mathrm{approx})}(\omega)$ approximates $\rho^{(\mathrm{true})}(\omega)$ within a few percent. On the other hand, a large error band  appears for $\Gamma=0.1$. The unphysical oscillation seen in Fig.~\ref{fig:mockbreitwigner_reconstruction_gammadep} is masked by the truncation error, and the error explodes toward $\omega\to 0$. 

Note that our approach to estimate the systematic error assumes only the scaling of $|\rho_l|$ and the triangle inequality. Therefore, it provides an upper bound of the truncation error. A more realistic error estimate may be obtained by further assuming a flat distribution of $\rho_l$ between $\pm A_\rho e^{-B_\rho l}$. A similar approach has been taken in the analysis of inclusive semileptonic decays for the amplitude of Chebyshev polynomials \cite{Kellermann:2025pzt,Kellermann:2026sgp}.

%
%
\begin{figure*}[tb]
\centering
\includegraphics[width=0.49\textwidth,bb=0 0 425 325,clip]{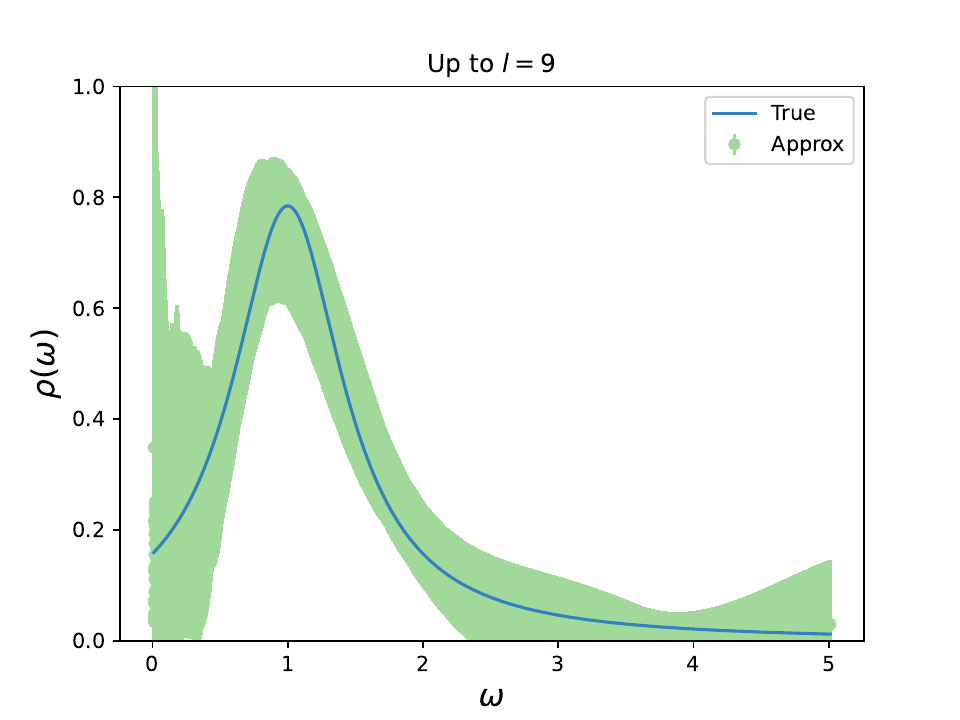}
\caption{
$\rho(\omega)$ (green square) with $\Gamma=0.5$ using up to $l=9$, compared with the true one (blue circle).
}
\label{fig:mockbreitwigner_reconstructionerr_ntr10}
\end{figure*}

In practice, due to statistical noise, the SVD components $\rho_l$ would not be obtained up to the limit of double precision. If instead determined only up to $l=9$, the error bound becomes much looser even for $\Gamma=0.5$ as shown in Fig.~\ref{fig:mockbreitwigner_reconstructionerr_ntr10}.
That is a motivation to consider the smeared spectrum, which is the subject of the following section.

\section{Smeared spectrum}
\label{sec:numerical_application}

We consider the smeared spectrum defined as a convolution integral with some kernel function $S_{\sigma_{\mathrm{smr}}}(\omega,\omega^\prime)$, which includes a parameter $\sigma_\mathrm{smr}$ that controls the size of the smearing. It can be expanded using the SVD basis as
\begin{align}
    \rho_{\sigma_{\mathrm{smr}}}(\omega)
    & =
    \int^\infty_0 d\omega^\prime\,
    S_{\sigma_{\mathrm{smr}}}(\omega,\omega^\prime)
    \rho(\omega)\nonumber\\
    \label{eq:smear_spec_approx}
    &\approx
    \sum_{l=0}^{N_\mathrm{tr}-1} \rho_l
    \left[
    \sum_{j=1}^{N_\omega}\Delta^\omega
    S_{\sigma_{\mathrm{smr}}}(\omega,\omega_j)
    V_l(\omega_j)
    \right] 
    =
    \sum_{l=0}^{N_\mathrm{tr}-1}\rho_l S_{\sigma_\mathrm{smr},l}(\omega).
\end{align}
The SVD components $S_{\sigma_\mathrm{smr},l}(\omega)$ are defined by the expression in the square bracket, while $\rho_l$ is given by the physical spectrum as discussed in the previous section. Therefore, this formula provides a separation of contributions: those of the physical spectrum $\rho_l$ and those of the smearing kernel $S_{\sigma_\mathrm{smr},l}(\omega)$.
Although the form of $S_{\sigma_\mathrm{smr}}(\omega,\omega^\prime)$ is arbitrary, for demonstration purposes, let us assume a specific form 
\begin{align}
    \label{eq:smear_kernel}
    S_{\sigma_\mathrm{smr}}(\omega,\omega^\prime)
    =
    \frac{1}{\mathcal{N}}
    \left[
    \frac{1}{\pi}
    \frac{2\sigma_\mathrm{smr}}{(\omega-\omega')^2+\sigma_\mathrm{smr}^2}.
    \right]
\end{align}
with a normalization factor $\mathcal{N}$.

\begin{figure*}[tb]
\centering
\includegraphics[width=0.49\textwidth,bb=0 0 425 325,clip]{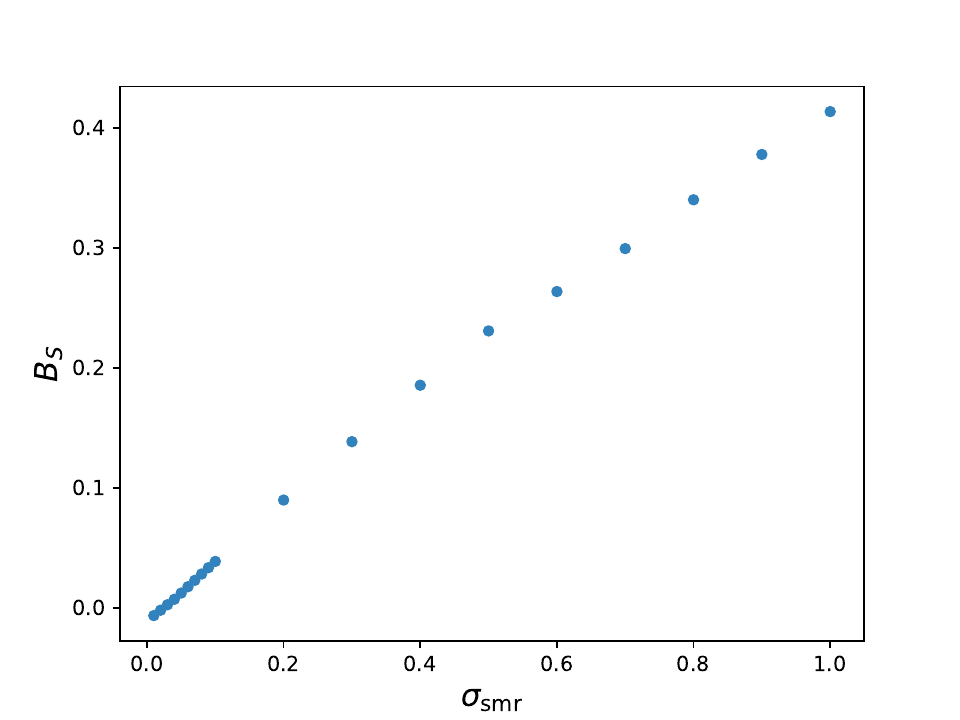}
\caption{
Exponent $B_S$ of the exponential falloff of $|S_{\sigma_\mathrm{smr},l}(\omega)|\leq A_S\mathrm{e}^{-B_S l}$. 
}
\label{fig:mockbreitwigner_smearfunc_power}
\end{figure*}

The truncation error of (\ref{eq:smear_spec_approx}) is controlled by how $|\rho_l|$ and $|S_{\sigma_\mathrm{smr},l}(\omega)|$ decrease for large $l$. As discussed in the previous section, the exponential falloff of $|\rho_l|$ depends on the functional form of the physical spectrum $\rho(\omega)$.
On the other hand, the exponential falloff of $|S_{\sigma_\mathrm{smr},l}(\omega)|$ is controlled by $\sigma_\mathrm{smr}$. In Fig.~\ref{fig:mockbreitwigner_smearfunc_power} we plot the exponent $B_S$ that characterizes the exponential form as $|S_{\sigma_\mathrm{smr},l}(\omega)|\leq A_S e^{-B_s l}$. As an example, we take $\omega=1.0$ but the result is similar for other $\omega$'s.

%
%

From Fig.~\ref{fig:mockbreitwigner_smearfunc_power} we observe that $B_S$ increases linearly with $\sigma_\mathrm{smr}$. Combined with the scaling of $|\rho_l|$, the truncation error of (\ref{eq:smear_spec_approx}) is controlled by the exponential form $e^{-(B_\rho+B_S)l}$, and the sum to $l\to\infty$ gives $A_\rho A_Se^{-(B_\rho+B_S)N_\mathrm{tr}}/(1-e^{-(B_\rho+B_S)})$. This explains how smearing suppresses the truncation error. When suppression by the spectrum itself $e^{-B_\rho l}$ is not sufficient, one can add an extra source of suppression by introducing $e^{-B_S l}$ at the cost of modifying the problem.

For illustration, we consider the physical spectrum $\rho(\omega)$ obeying the Breit-Wigner distribution (\ref{eq:breit_wigner}) with $\omega_0=1.0$ and $\Gamma=0.5$. We construct $\rho_l$ from $C(t)$ with (\ref{eq:spec_rep}) and (\ref{eq:rhomode_t}) using SVD. With a truncation at $N_\mathrm{tr}=9$, we obtain the original estimate of the physical spectrum shown in Fig.~\ref{fig:mockbreitwigner_reconstructionerr_ntr10} together with the error bound. The scaling of $|\rho_l|$ gives $B_\rho$ = 0.21.

%
%
\begin{figure*}[tb]
\centering
\includegraphics[width=0.49\textwidth,bb=0 0 425 325,clip]{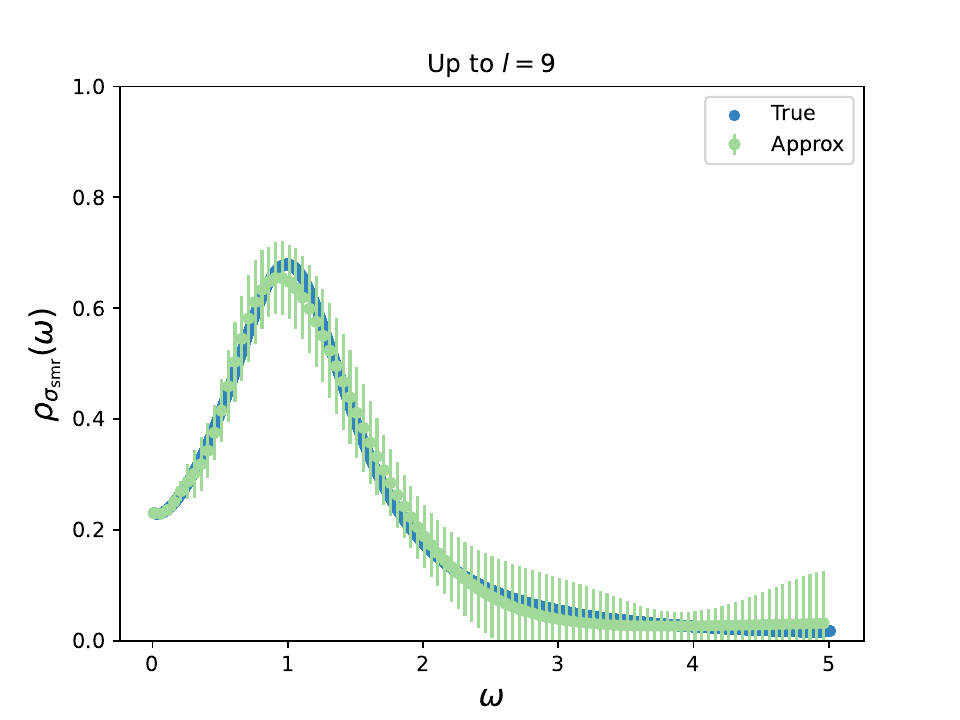}
 \includegraphics[width=0.49\textwidth,bb=0 0 425 325,clip]{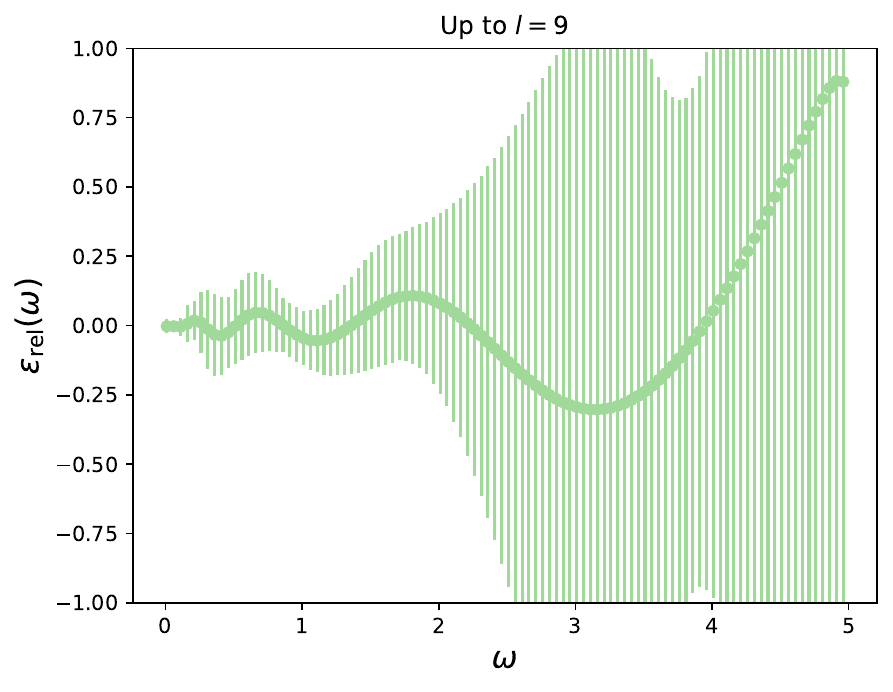}
\includegraphics[width=0.49\textwidth,bb=0 0 425 325,clip]{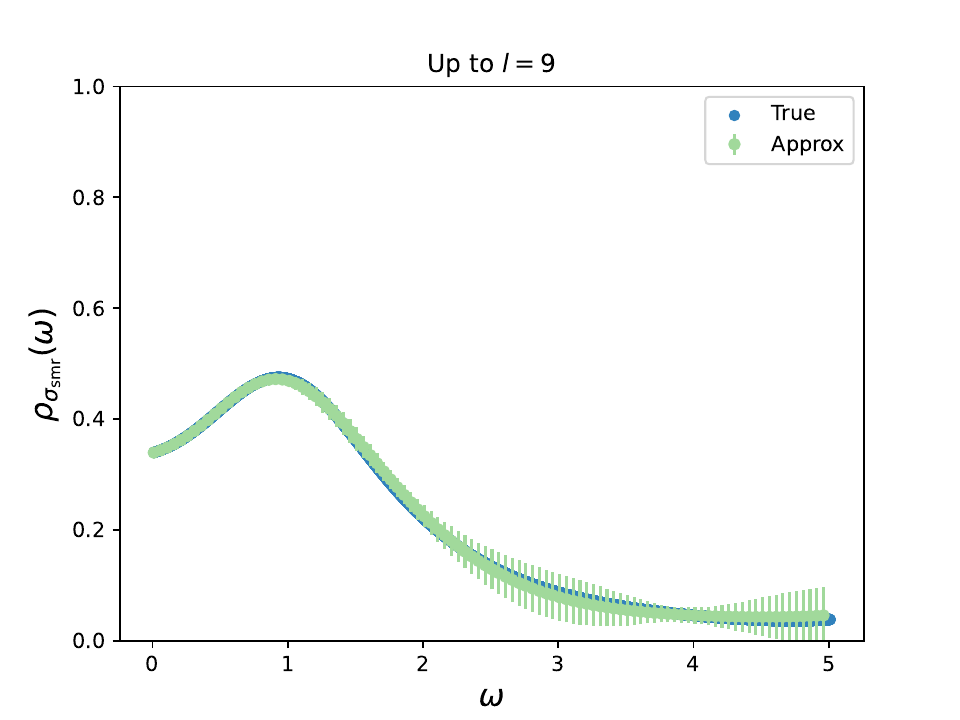}
 \includegraphics[width=0.49\textwidth,bb=0 0 435 325,clip]{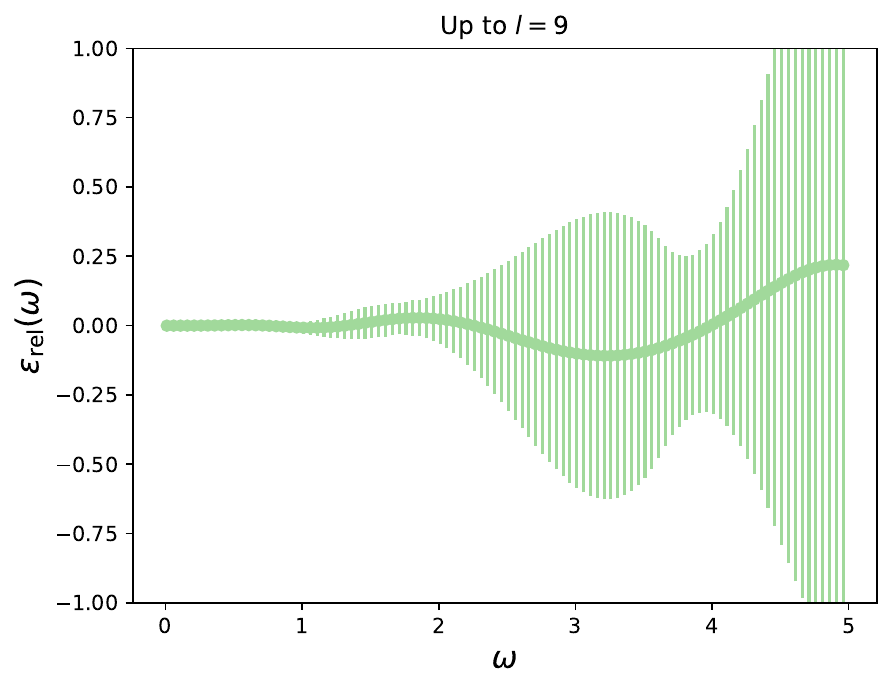}
\includegraphics[width=0.49\textwidth,bb=0 0 425 325,clip]{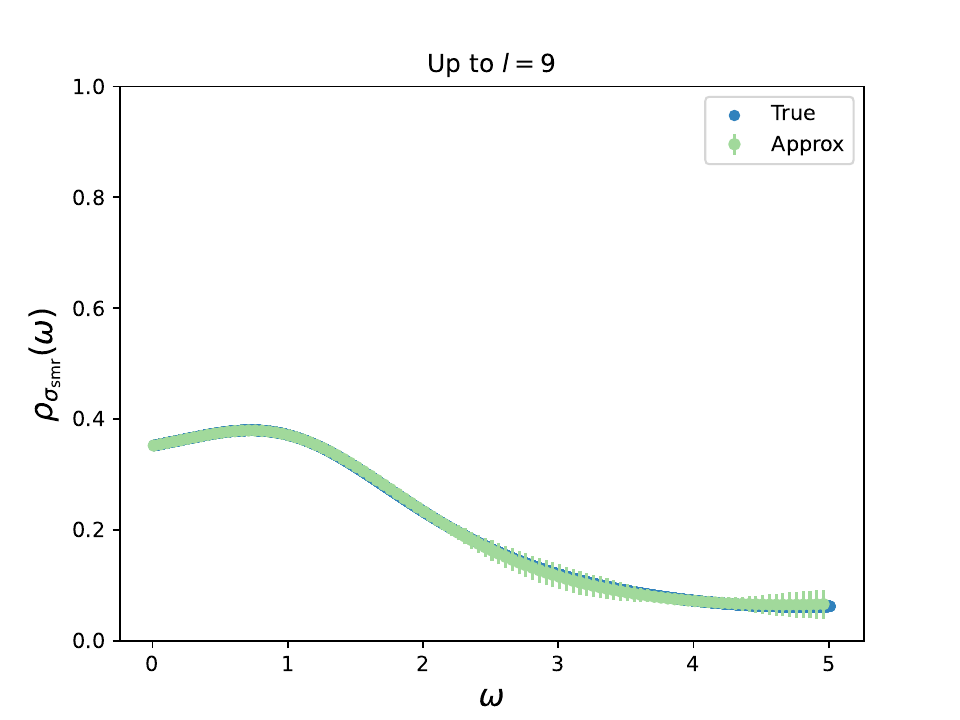}
 \includegraphics[width=0.49\textwidth,bb=0 0 435 325,clip]{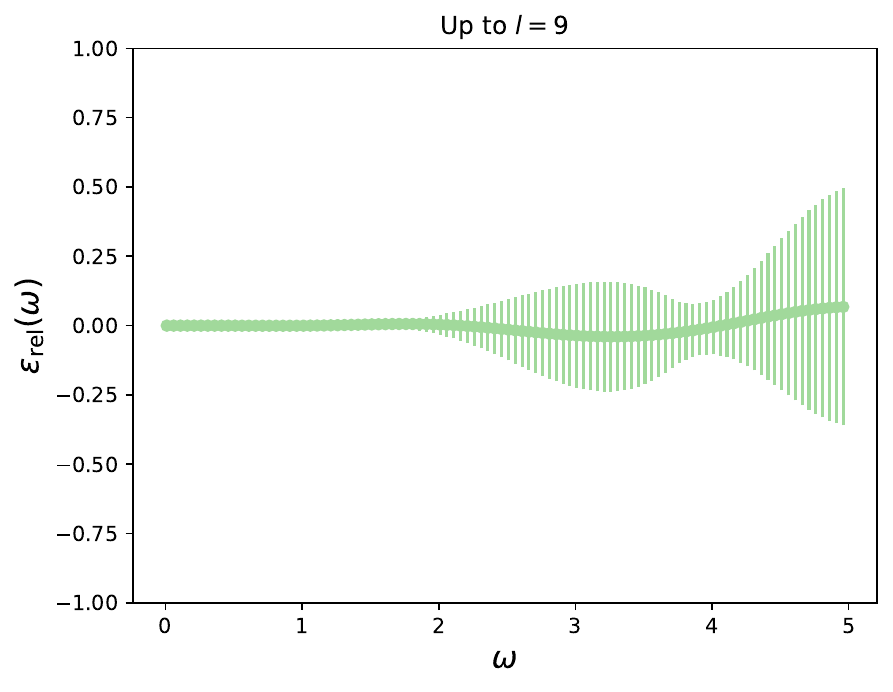}
\caption{
$\rho^\mathrm{(approx)}(\omega)$ (green square) with truncation error, and the true function (blue circle).
Each plot is obtained with $\sigma_\mathrm{smr}=0.1$ (top), $0.5$ (middle) and $1.0$ (bottom). Their relative error is shown in the right panels.
}
\label{fig:mockbreitwigner_reconstructionerr_smear}
\end{figure*}

Fig.~\ref{fig:mockbreitwigner_reconstructionerr_smear} shows the smeared spectrum $\rho_{\sigma_\mathrm{smr}}(\omega)$ with $\sigma_\mathrm{smr}=0.1$ (top), $0.5$ (middle), and $1.0$ (bottom). 
The relative error to the true function is shown in the right panels. As anticipated, the error is significantly reduced with smearing. 

\begin{figure*}[tb]
\centering
\includegraphics[width=0.49\textwidth,bb=0 0 425 325,clip]{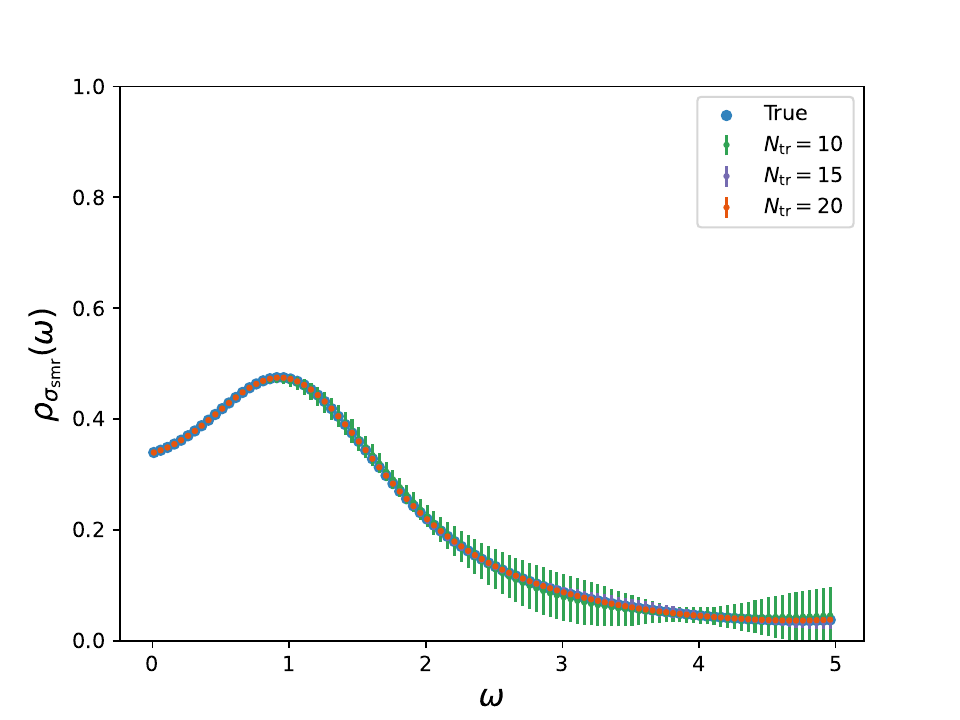}
\includegraphics[width=0.49\textwidth,bb=0 0 435 325,clip]{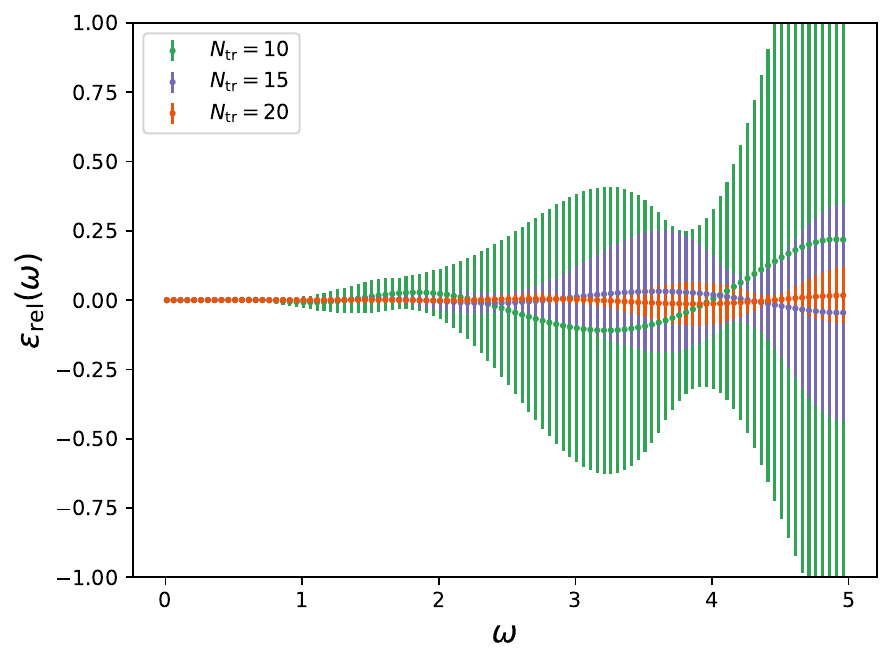}
\caption{
$\rho^{(\mathrm{approx})}(\omega)$ for $\sigma_\mathrm{smr}=0.5$ obtained with $N_\mathrm{tr}=10$ (green), $15$ (violet), $20$ (orange), and the true one (blue circle).
}
\label{fig:mockbreitwigner_reconstructionerr_ntrdep_smear}
\end{figure*}

The approximation improves with larger $N_\mathrm{tr}$. Fig.~\ref{fig:mockbreitwigner_reconstructionerr_ntrdep_smear} shows $\rho_{\sigma_\mathrm{smr}}^{(\mathrm{approx})}(\omega)$ for $\sigma_\mathrm{smr}=0.5$ obtained with $N_\mathrm{tr}=10$ (green), $15$ (violet), $20$ (orange). With this smearing width, the truncation error is suppressed to a few \% level if one can include $N_\mathrm{tr}=20$ terms.

%
%
%

\section{Summary}
\label{sec:summary}

Reconstruction of spectral function from Euclidean lattice correlators remains a difficult problem. As a linear system, it corresponds to the inversion of a matrix made of $e^{-\omega t}$ for a discrete set of $t$ and $\omega$. The inversion is numerically ill-defined when the matrix has extremely small eigenvalues. In order to regulate the problem, we introduce the SVD of the kernel $e^{-\omega t}$ as a mathematical tool.

The SVD defines singular values and associated basis vectors. Smaller singular values have more oscillatory basis vectors, corresponding to more detailed structure of the spectral function. For inversion, only terms that have relatively large singular values can be kept, while terms with too small singular values suffer from enhanced noise. By truncating the SVD basis, one can obtain a numerically stable estimate of the spectral function, albeit with sometimes significant truncation errors.

The truncation error can be estimated transparently within the SVD approach, under the assumption that the SVD components of the spectral function decrease exponentially for smaller singular values. A large error is expected when the resulting spectral function contains narrow peaks and the SVD components does not decay rapidly. The smeared spectral function could be considered to tame the large truncation error; the remaining error can be rigorously estimated again using the exponential suppression of the SVD components of the smearing kernel. A similar but slightly different error estimate is available with the Chebyshev approximation approach \cite{Kellermann:2025pzt}.

The SVD basis is smoothly connected to the Mellin basis discussed in \cite{Bruno:2024fqc} in the limit of vanishing temporal lattice spacing and infinite temporal separation. Therefore, SVD provides a natural means for implementing the Mellin transform for the lattice data obtained at finite temporal extent. It would be interesting to define smeared physical quantities that can be accessed by experiments using this setup so that direct comparison with the lattice calculation is possible.

Overall, the SVD approach provides a transparent basis for understanding the problem in the reconstruction of the spectral function and for estimating systematic errors. The core of the problem lies in the fact that the amount of information carried by the Euclidean correlator is inherently limited, and it cannot be circumvented by any clever method without introducing extra assumptions. 

The SVD basis can be used to evaluate the smeared spectrum or any weighted integrals of the spectral function, such as the inclusive decay rate of heavy hadrons \cite{Gambino:2020crt}. It would provide equally effective methods as alternatives, {\it e.g.} HLT or Chebyshev polynomials. The advantage of the SVD approach is that it allows a rigorous estimate of the truncation error. In this respect, a comparison to other approaches would be interesting. The limit of vanishing smearing width is hard to reach unless some additional assumption is introduced.

There would be a number of other numerical applications of the method. In addition to inclusive decay rates and associated moments, the filtering of Euclidean correlators may open up a variety of new applications \cite{Bruno:2020kyl}. Reconstruction of Minkowski transition amplitudes is another area of interesting new possibilities \cite{Bulava:2019kbi,Frezzotti:2023nun}.  

With finite temperature, the kernel function must be modified, but the same strategy can be applied. An interesting application would be the QCD transport coefficients, including the shear viscosity that characterizes the property of quark-gluon plasma \cite{Hosoya:1983id, Wang:1995qg, Meyer:2007ic}. The problem of taking the limit of vanishing smearing width remains also in this case. 

\begin{acknowledgments}
R.T. is supported in part by JSPS KAKENHI Grant Number JP25KJ0404.
The works of S.H. are supported in part by JSPS KAKENHI Grant Numbers 22H00138.
\end{acknowledgments}
\appendix

\clearpage
\section{Singular values and basis vectors}
\label{app:the_scaling_properties}

\begin{figure*}[tb]
\centering
\includegraphics[width=0.49\textwidth,bb=0 0 425 305,clip]{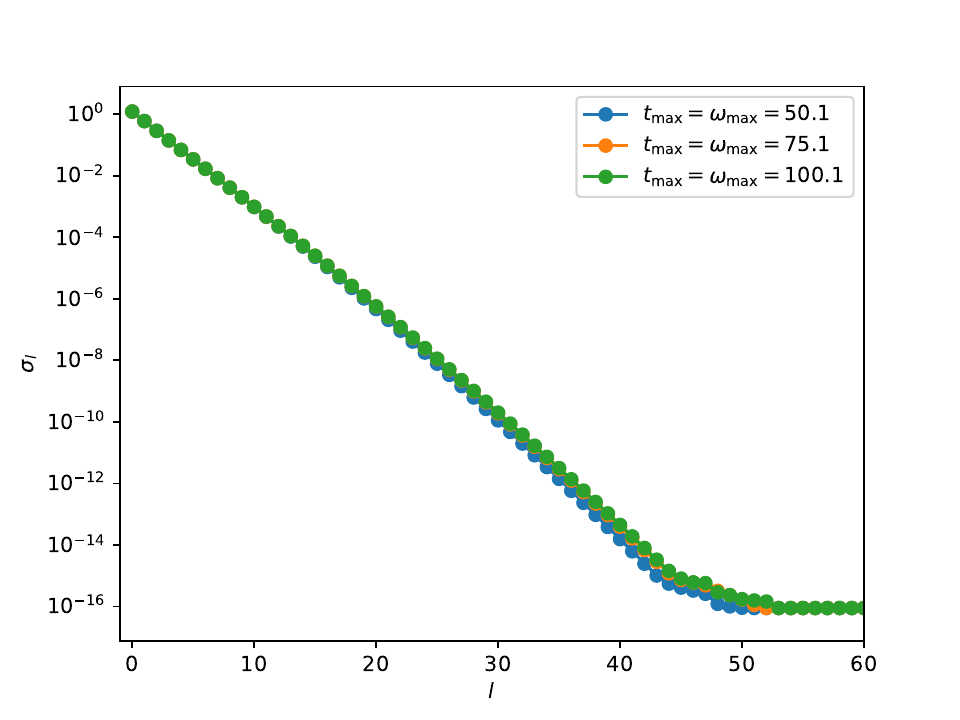}
\caption{
Singular values $\sigma_l$ for $t_\mathrm{max}=\omega_\mathrm{max}=10.1$ (blue), $50.1$ (orange), and $100.1$ (green). The other parameters are 
$t_\mathrm{min}=\omega_\mathrm{min}=0.1$ and  $\Delta=0.05$. 
}
\label{fig:symmetric_rangedep_singularvalue}
\end{figure*}

Singular values $\sigma_l$ and basis vector $\tilde{U}_l$ and $\tilde{V}_l$ defined for $\tilde{L}$ with $t_\mathrm{min}=\omega_\mathrm{min}$, $\Delta^t=\Delta^\omega\equiv\Delta$ and $N_t=N_\omega$ are summarized. Here, we take $t_\mathrm{min}=\omega_\mathrm{min}=0.1$ and $\Delta=0.05$. Fig.~\ref{fig:symmetric_rangedep_singularvalue} shows $\sigma_l$ for $t_\mathrm{max}=\omega_\mathrm{max}$ = 50.1, 75.1, 100.1. The dependence on $t_\mathrm{max}=\omega_\mathrm{max}$ is insignificant; a slight difference is found for the smallest $t_\mathrm{max}=\omega_\mathrm{max}$ = 50.1.

%
%

\begin{figure*}[tb]
\includegraphics[width=0.49\textwidth,bb=0 0 425 305,clip]{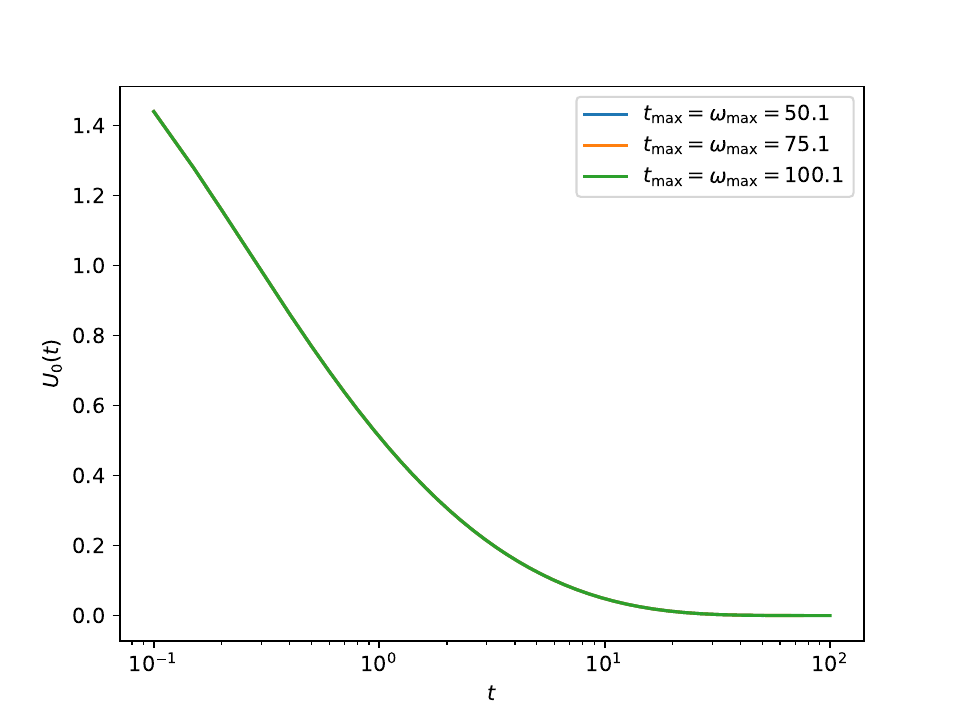}
\includegraphics[width=0.49\textwidth,bb=0 0 425 305,clip]{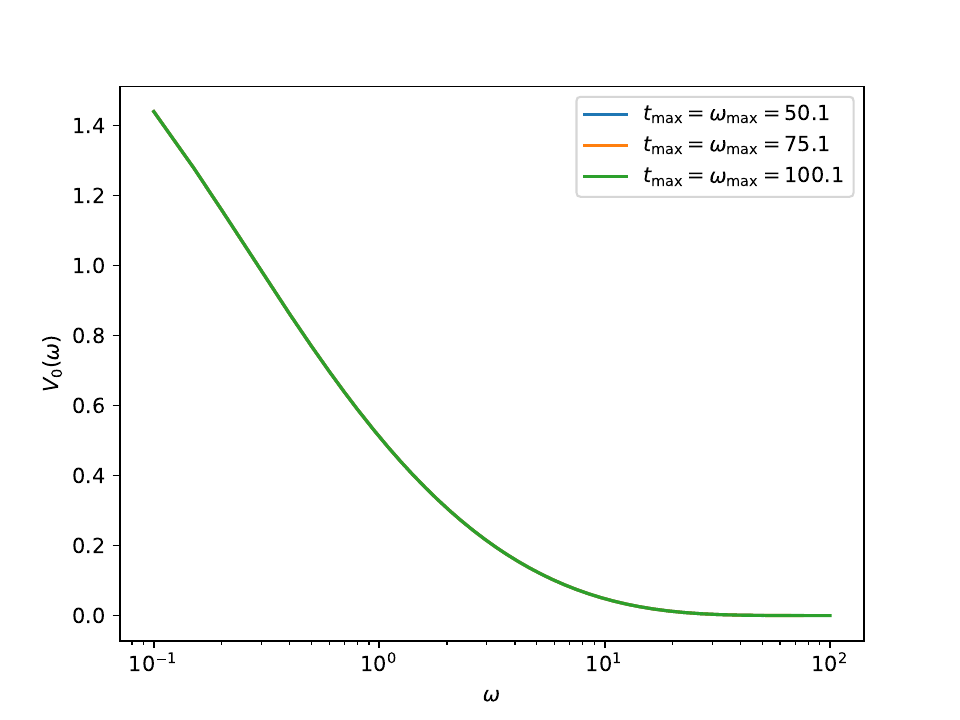}
\includegraphics[width=0.49\textwidth,bb=0 0 425 305,clip]{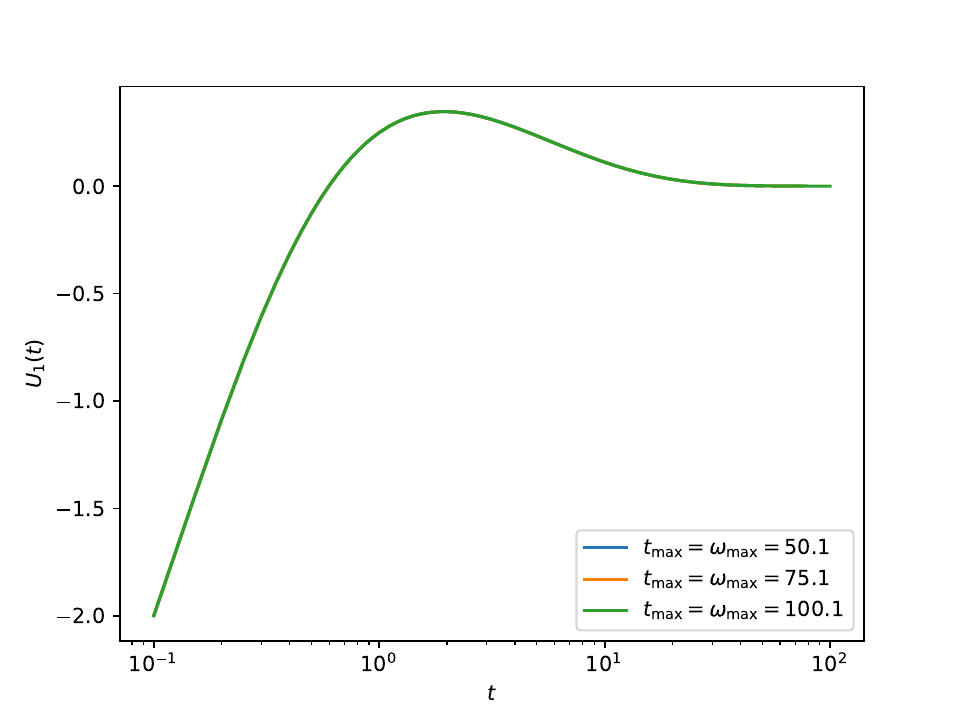}
\includegraphics[width=0.49\textwidth,bb=0 0 425 305,clip]{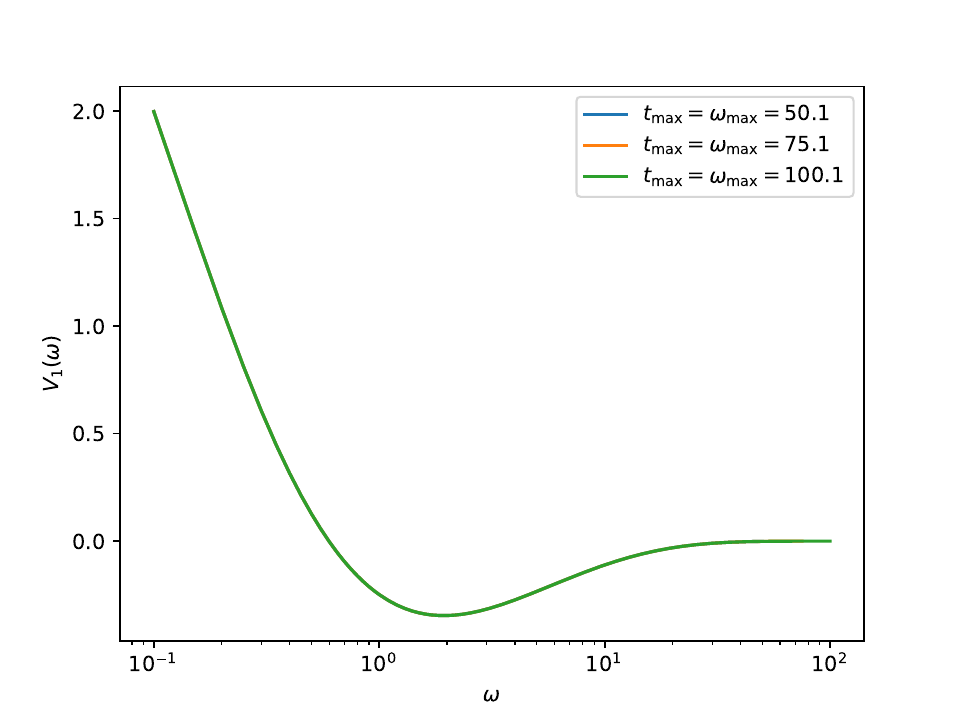}
\includegraphics[width=0.49\textwidth,bb=0 0 425 305,clip]{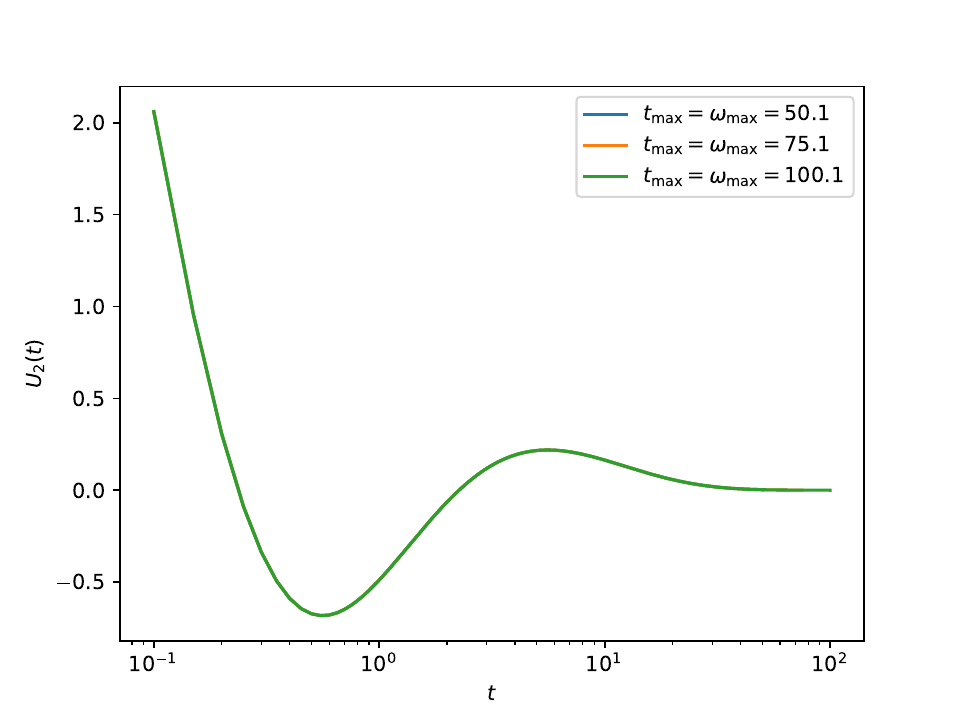}
\includegraphics[width=0.49\textwidth,bb=0 0 425 305,clip]{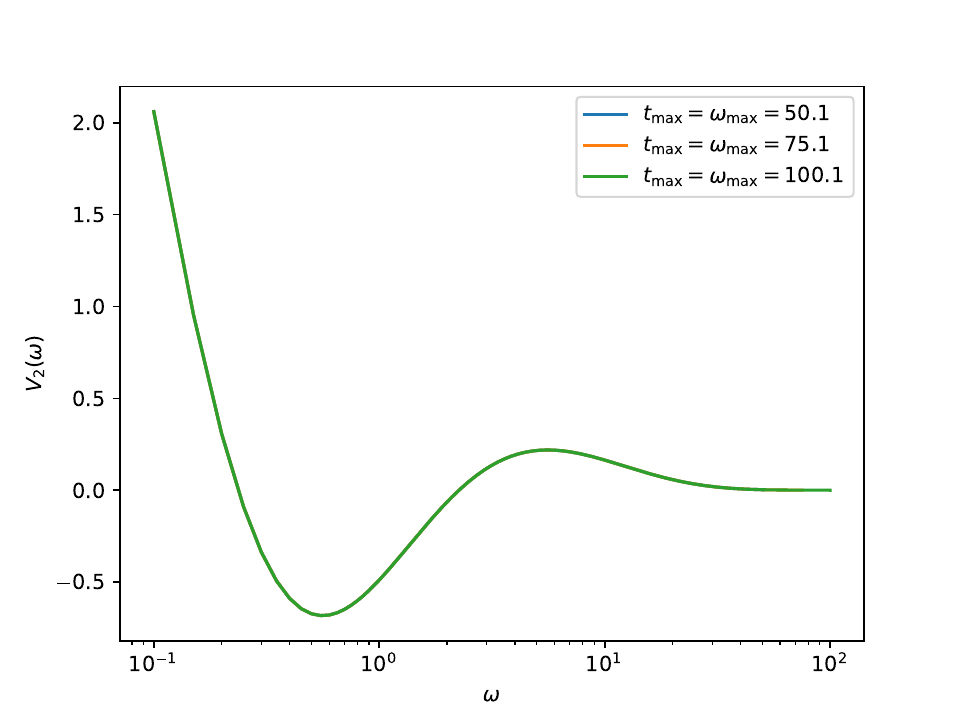}
\caption{
$U_l(t_i)$ (left) and $V_l(\omega_j)$ (right) for $l$ = 0, 1, 2. Plots are with $t_\mathrm{max}=\omega_\mathrm{max}=50.1$ (blue), $75.1$ (orange), and $100.1$ (green), while $t_\mathrm{min}=\omega_\mathrm{min}=0.1$ and $\Delta=0.05$ are fixed.
}
\label{fig:symmetric_rangedep_basisfunc_low}
\end{figure*}

\begin{figure*}[tb]
\includegraphics[width=0.49\textwidth,bb=0 0 425 305,clip]{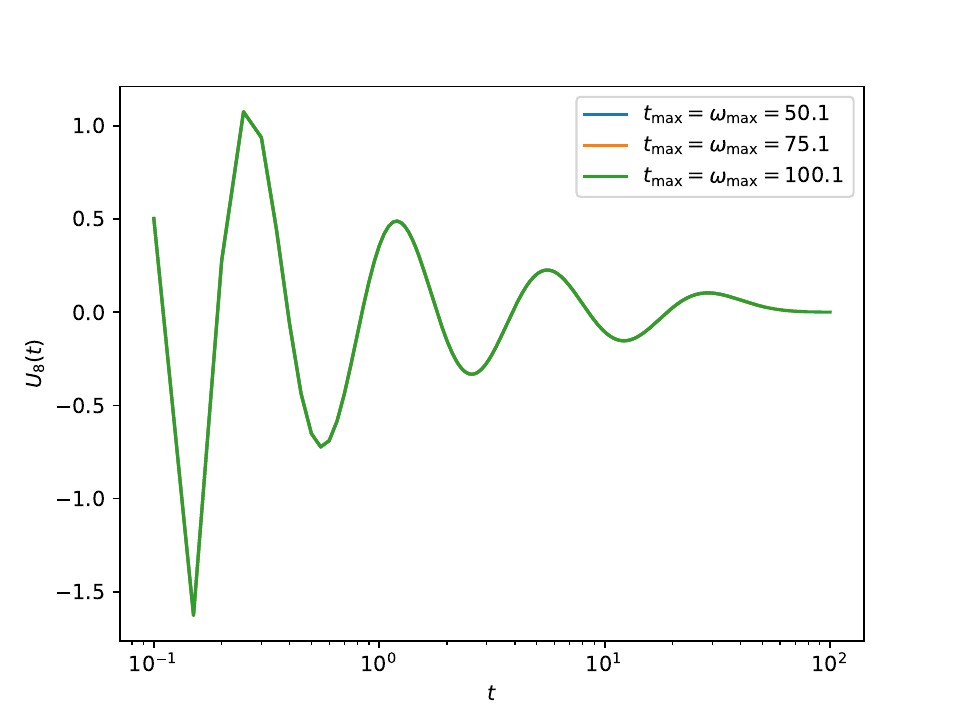}
\includegraphics[width=0.49\textwidth,bb=0 0 425 305,clip]{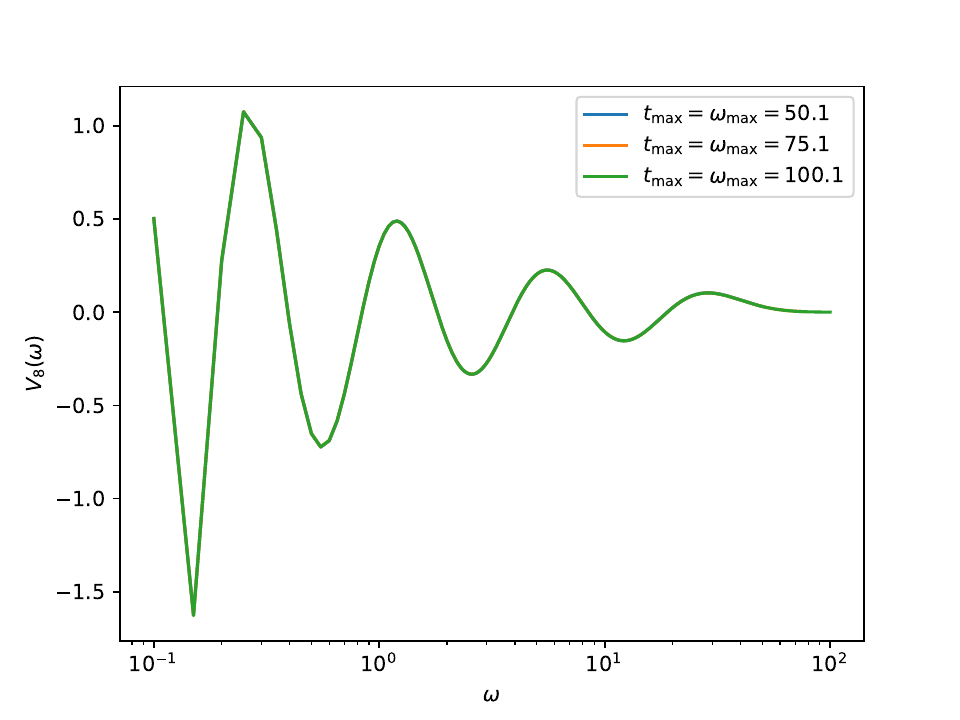}
\includegraphics[width=0.49\textwidth,bb=0 0 425 305,clip]{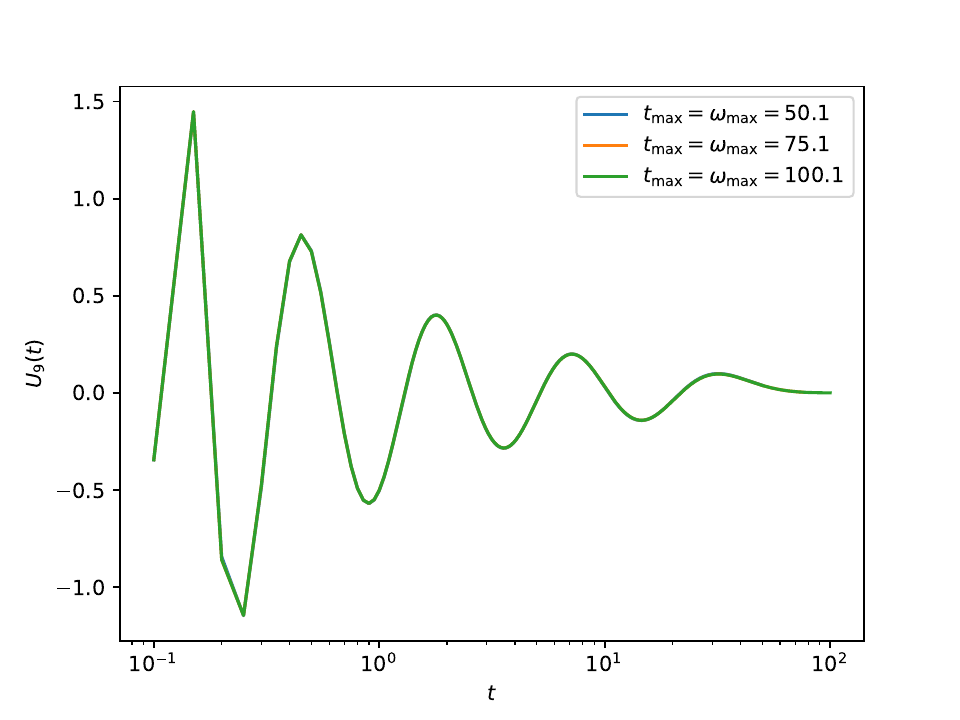}
\includegraphics[width=0.49\textwidth,bb=0 0 425 305,clip]{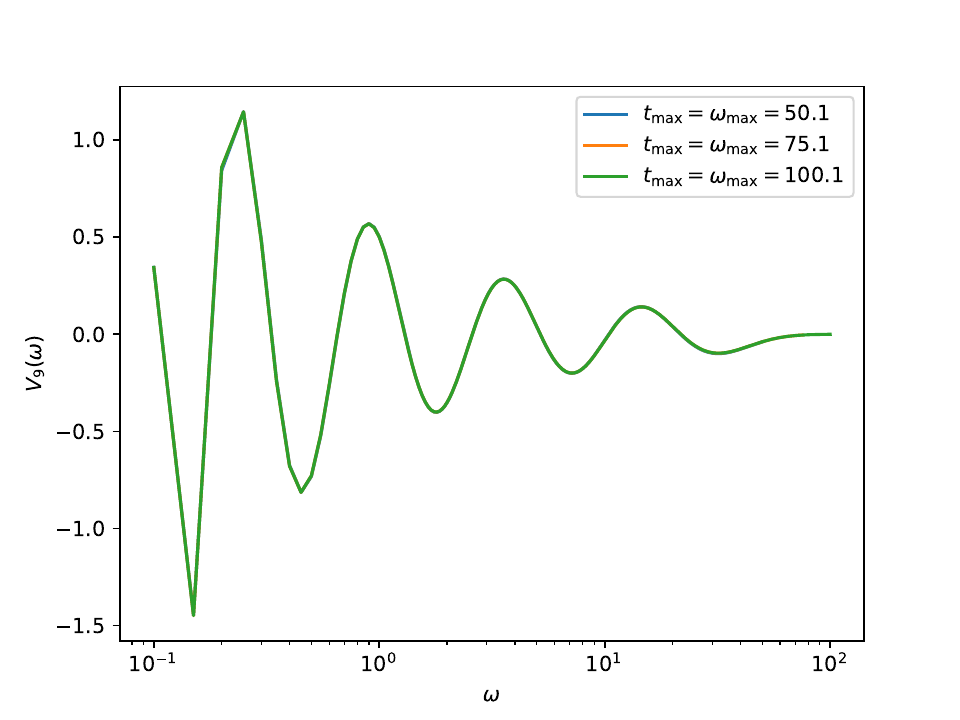}
\includegraphics[width=0.49\textwidth,bb=0 0 425 305,clip]{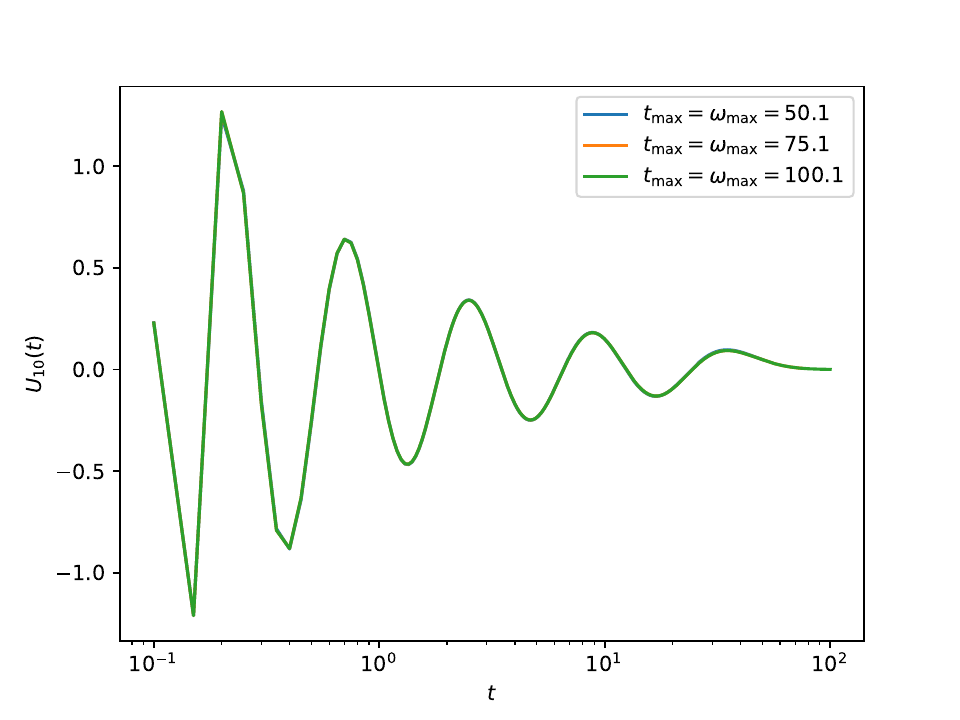}
\includegraphics[width=0.49\textwidth,bb=0 0 425 305,clip]{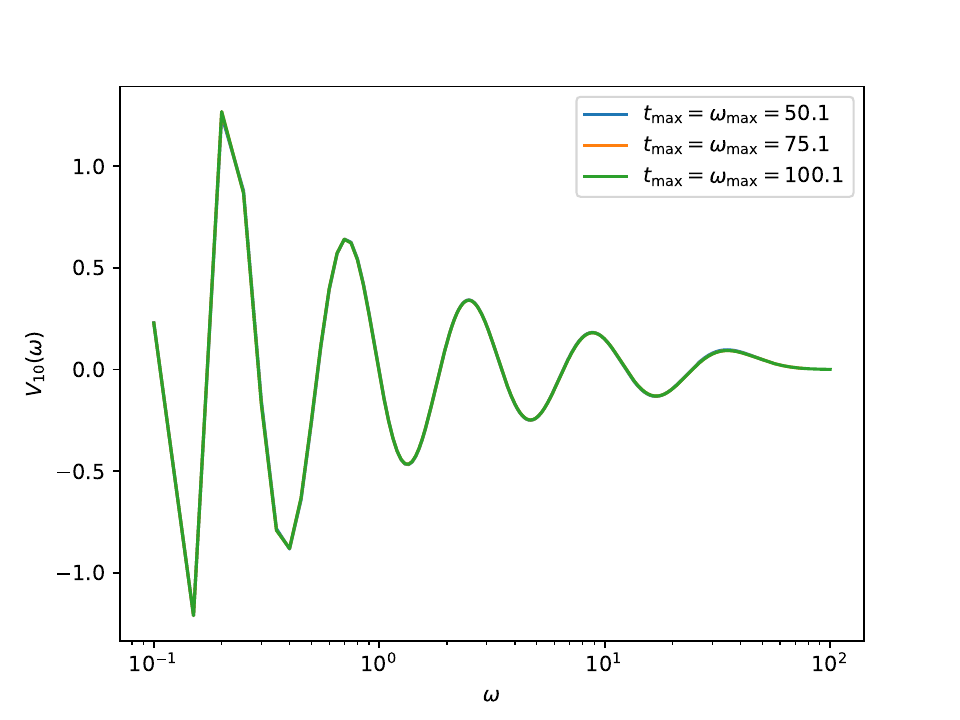}
\caption{
Same as Fig.~\ref{fig:symmetric_rangedep_basisfunc_low} but for
$l$ = 8, 9, 10.
}
\label{fig:symmetric_rangedep_basisfunc_int}
\end{figure*}

\begin{figure*}[tb]
\includegraphics[width=0.49\textwidth,bb=0 0 425 305,clip]{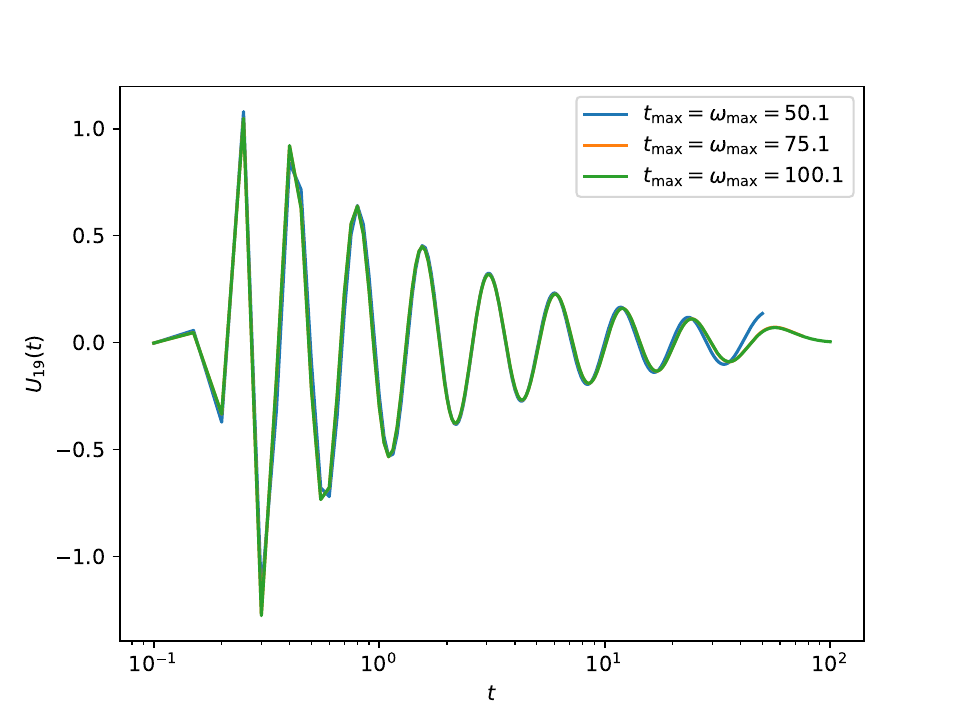}
\includegraphics[width=0.49\textwidth,bb=0 0 425 305,clip]{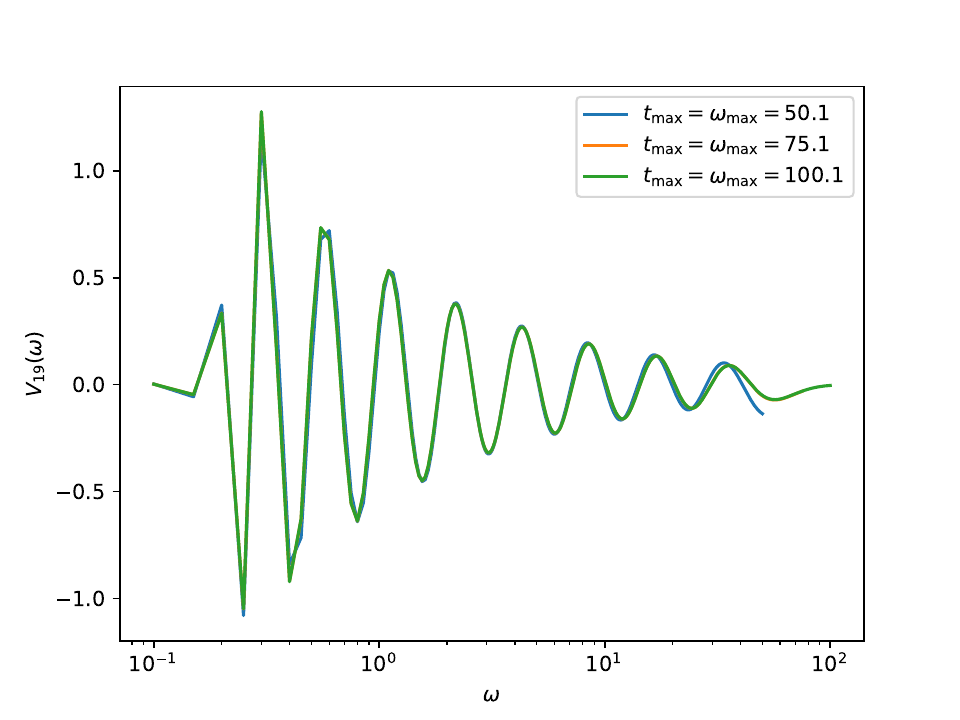}
\includegraphics[width=0.49\textwidth,bb=0 0 425 305,clip]{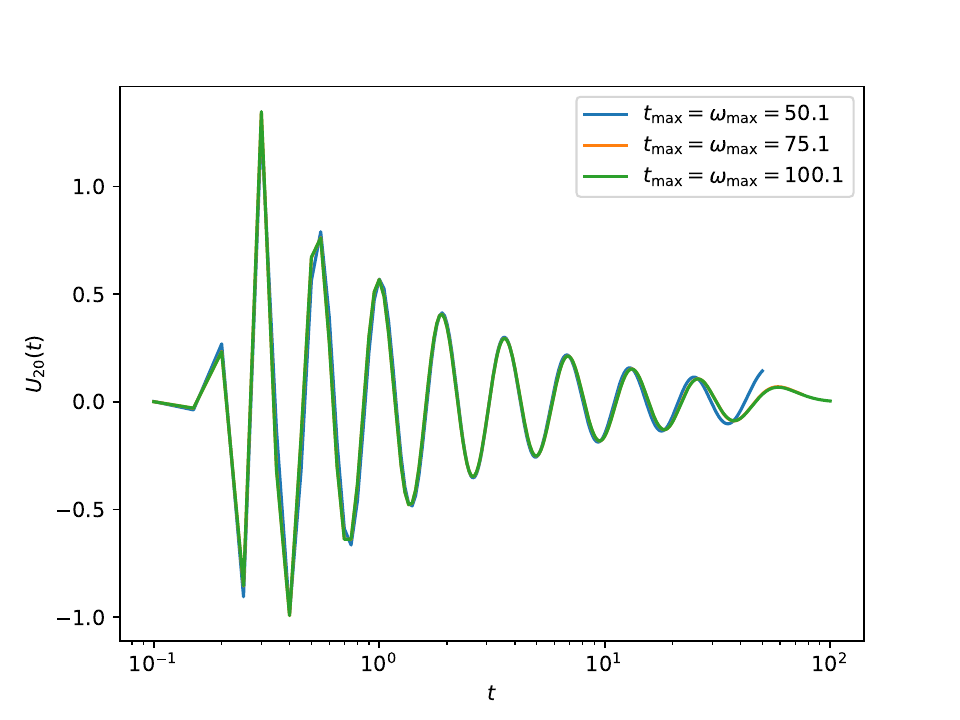}
\includegraphics[width=0.49\textwidth,bb=0 0 425 305,clip]{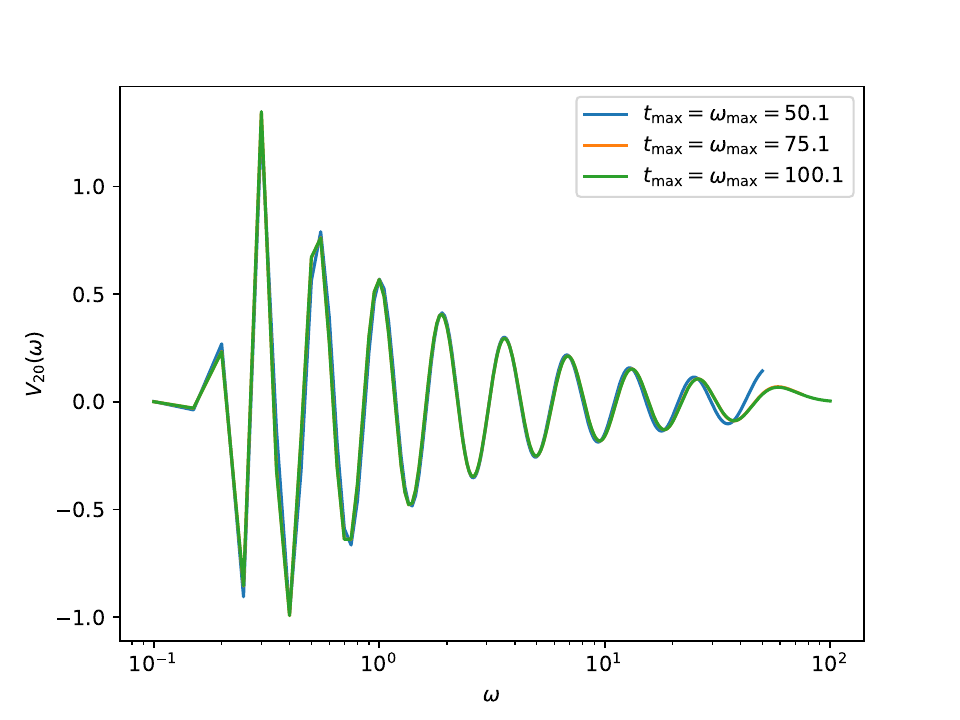}
\includegraphics[width=0.49\textwidth,bb=0 0 425 305,clip]{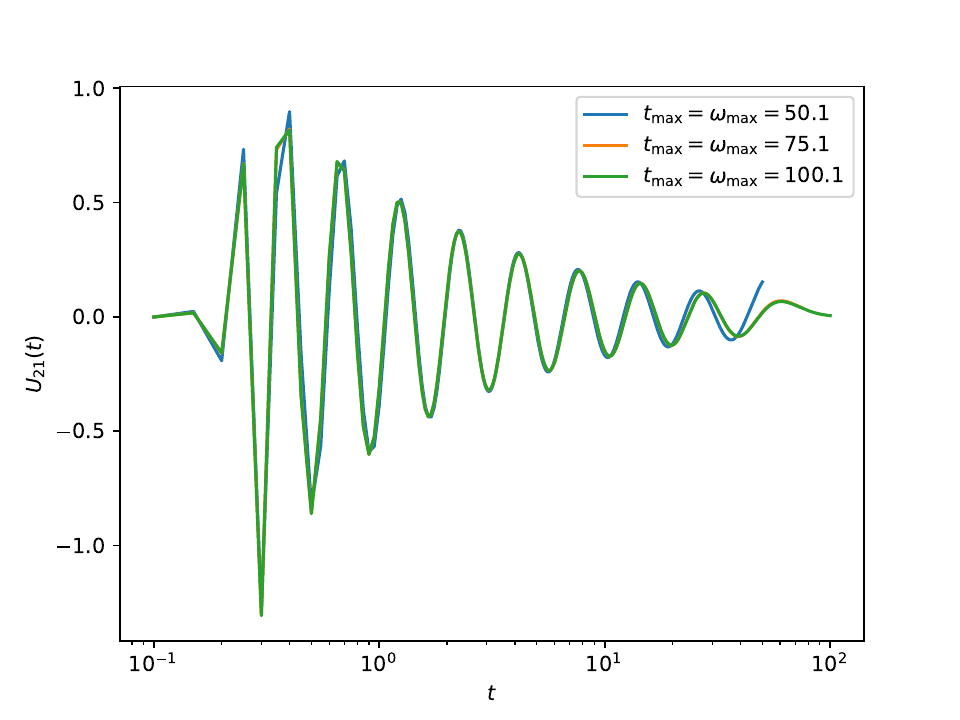}
\includegraphics[width=0.49\textwidth,bb=0 0 425 305,clip]{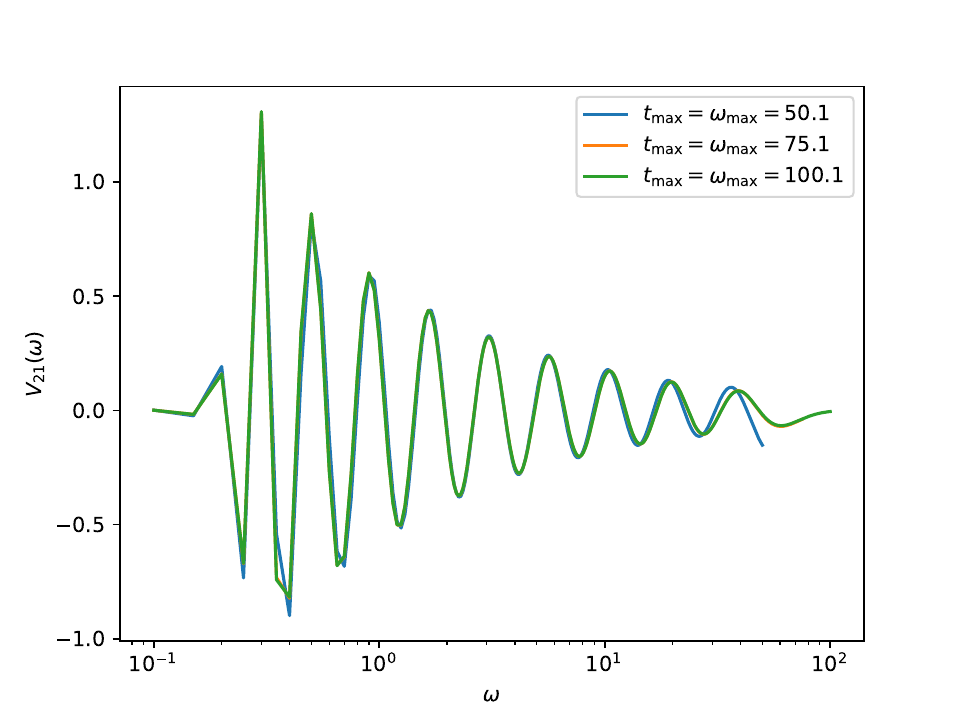}
\caption{
Same as Fig.~\ref{fig:symmetric_rangedep_basisfunc_low} but for 
$l$ = 19, 20, 21.
}
\label{fig:symmetric_rangedep_basisfunc_high}
\end{figure*}

Figs.~\ref{fig:symmetric_rangedep_basisfunc_low} and \ref{fig:symmetric_rangedep_basisfunc_int} show the basis functions  $U_l(t_i)=(\tilde{U})_{li}/\sqrt{\Delta}$ and  $V_l(\omega_j)=(\tilde{V^t})_{lj}/\sqrt{\Delta}$ for $l$ = 0, 1, 2 and $l$ = 8, 9, 10, respectively. We do not observe any significant dependence on $t_\mathrm{max}=\omega_\mathrm{max}$. 
Similar plots for $l$ = 19, 20, 21 are shown in Fig.~\ref{fig:symmetric_rangedep_basisfunc_high}.
%
%
%

%
%
%
%
%
%

\bibliography{SpectralFunctionWithSVD/main}
\bibliographystyle{man-apsrev}
 
\end{document}